# Reversible Interchain Associations Dominate Plateau Elasticity in Conjugated Polymers

Juncheng Zheng,[1] Xiaobing Zuo,[2] Xiaodan Gu,[3,*] and Shiwang Cheng[1,*]

[1]Department of Chemical Engineering and Materials Science, Michigan State University, East Lansing, Michigan 48824, United States
[2]X-ray Science Division, Argonne National Laboratory, Lemont, Illinois 60439, United States
[3]School of Polymer Science and Engineering, Center for Optoelectronic Materials and Devices, The University of Southern Mississippi, Hattiesburg, Mississippi 39406, United States

## Abstract

The modulus plateau is widely regarded as a hallmark of topological entanglement in homopolymer melts and has been used to identify entanglement in semiflexible conjugated polymers. Using regiorandom poly(3-alkylthiophenes) as a model system, we show that their plateau elasticity relaxes through an Arrhenius process decoupled from segmental and terminal dynamics, reorganizes persistently under mechanical perturbation, and varies systematically with side-chain length. These behaviors are hard to reconcile with a purely topological-entanglement picture and instead support a transient network dominated by reversible interchain associations, consistent with $\pi-\pi$ interactions between conjugated backbone segments. The results thus demonstrate that plateau elasticity alone does not uniquely identify topological entanglement in conjugated polymers. More broadly, they identify the strength and lifetime of interchain association as molecular design parameters linking backbone packing to stress relaxation, viscoelastic properties, and processability in conjugated polymers, complementing their established role in electronic structure and charge transport.

* Corresponding Authors. Email Address: xiaodan.gu@usm.edu (X.G) and chengsh9@msu.edu (S.C.)

## INTRODUCTION

Conjugated polymers underpin a rapidly expanding range of technologies, including flexible and wearable electronics, bioelectronics, and sensing,[1-4] yet predictive molecular design remains limited by an incomplete understanding of the underlying polymer physics. A central issue is topological entanglement,[5-9] which in conventional linear polymers connects molecular weight and chain architecture to long-time relaxation, viscoelastic properties, and mechanical responses.[10-12] Conjugated polymers, however, combine semirigid backbones with flexible side chains and strong interchain associations, introducing molecular packing and relaxation behaviors distinct from that of conventional flexible polymers.[13-15] Establishing whether—and under what molecular conditions—these polymers form topological entanglements is therefore essential for developing molecular design rules linking conjugated polymer chemistry to their mechanical properties and processing behavior.[16, 17]

Recent studies have reported pronounced modulus plateaus in linear conjugated polymers and interpreted them as evidence of topological entanglement.[14-17] This interpretation has further motivated the use of conjugated polymers as model systems for developing broader descriptions of entanglement in semiflexible chains.[16, 17] The validity of these conclusions therefore depends critically on whether the modulus plateau uniquely reflects topological constraints. A plateau, however, establishes only the presence of long-lived elastic constraints, not their microscopic origin.[18-22] If plateau elasticity in conjugated polymers is not uniquely topological, both the assignment of entanglement in these materials and their broader use as model systems for semiflexible entanglement would require re-examination. This distinction is particularly important because neighboring conjugated backbone segments can form reversible interchain associations that could provide an alternative source of long-lived elasticity. From this perspective, distinguishing topological entanglement from interaction-mediated constraints for network elasticity is therefore a central unresolved problem in conjugated polymers.

Here we test directly whether the modulus plateau in conjugated polymers uniquely signifies topological entanglement. Using amorphous regiorandom poly(3-alkylthiophenes) (P3ATs) as a model system (**Figures 1a-1b** and **Figures S1-S4**), we combine creep recovery, temperature-dependent relaxation, and controlled mechanical perturbations to probe the network elasticity, dynamics, and mechanical stability. We find that the plateau-forming network extends across multiple chains, relaxes through dynamics decoupled from segmental and terminal motion, and undergoes long-lived reorganization upon mechanical deformation. These results reveal a form of plateau elasticity in P3ATs that is not captured by a conventional topological-entanglement picture and point to reversible interchain associations as an additional source of long-lived network constraints. This distinction provides a basis for reassessing how modulus plateaus are interpreted in conjugated polymers.

## RESULTS

### Distinct elastic recovery in conjugated polymers

Small- and wide-angle X-ray scattering (SAXS/WAXS) are employed to characterize the structure of regiorandom poly(3-hexylthiophene-2,5-diyl) (P3HT). No crystalline or liquid crystalline ordering is observed from 298 K to 533 K (**Figure 1a**). At the same time, differential scanning calorimetry (DSC) exhibits a glass transition step at $T_g \approx 273\ K$ and no additional thermal transitions at higher temperatures upto 533 K (**Figure 1b**). These results suggest the absence of crystalline or liquid crystalline ordering in the regiorandom P3HT, consistent with previous studies.[23, 24] From the size-exclusion chromatography (SEC) measurements (**Figure 1c,** details see **MATERIALS AND METHODS**), one gets its weight-average molecular weight, $M_w^{P3HT} \approx 21\ kg/mol$, and polydispersity, Đ = 2.4.

**Figure 1d** compares the linear viscoelastic spectra of this regiorandom P3HT and an entangled polystyrene (PS, $M_w^{PS} = 172\ kg/mol$ and Đ = 2.2, **Figure S2b**) used here as an entangled polymer reference. Small amplitude oscillatory shear (SAOS) spectra at different temperatures were shifted to reference temperatures of $T_{ref} =\ 333\ K$ for P3HT and $T_{ref} =\ 423\ K$ for PS. The corresponding dynamic shift factors, $a_T^S$, and modulus shift factors, $b_T$, are summarized in **Figure S5**. P3HT exhibits a qualitatively similar viscoelastic spectrum, including a pronounced intermediate-frequency modulus plateau of comparable width (around 3 decades) to that of PS, despite their difference in plateau moduli, $G_N$. The similarity in their linear viscoelastic properties has largely motivated the previous assignment of the P3HT plateau to topological entanglement.[6, 15, 16] However, a detailed analysis on the modulus plateau length leads to numbers of entanglement per chain, $Z \approx 8$ , for this PS. If the P3HT has a similar $Z \approx 8$ as the PS, one would anticipate a plateau modulus, $G_N \approx 1.2\ MPa$, that is $\sim 3$ times higher than experiments (**Figure 1d**). This seemingly discrepancy between the molecular-weight-based entanglement estimates and the experimentally observed plateau behavior challenges the interpretation of the modulus plateau in P3HT as arising from topological entanglements.

On the other hand, the PS spectra superpose over the full frequency range at different temperature, consistent with its thermorheological simplicity.[25, 26] In contrast, P3HT exhibits systematic deviations in the linear viscoelastic spectra at the plateau regime.[15] For instance, the corresponding loss factor, $tan\delta = G''/G'$, changes shape with testing temperature (inset of **Figure 1d**), implying a potential breakdown of the time-temperature superposition (TTS) of the P3HT at the intermediate frequencies.

To probe the underlying elastic response over broader times, we performed creep and creep-recovery measurements, extending the accessible window from approximately $10^{-2} - 10^{2}\ s$ in SAOS to $10^{-2} - 10^{5}\ s$.[27, 28] **Figure 1e** shows representative measurements for P3HT at $T = 318\ K$ and PS at $T = 383\ K$ under $\sigma_c = 1{,}000\ Pa$. This stress corresponds to nominal plateau strains, $\sigma_c/G_N$, of only ~0.2% for P3HT and ~0.7% for PS, a deformation range that would ordinarily be expected to lie within the linear-response regime.

For both polymers, the creep compliance, $J(t) = \gamma(t)/\sigma_c$, develops an intermediate-time plateau before entering terminal flow, $J(t){\sim}t$, where $\gamma(t)$ is the creep strain. The zero-shear viscosity is obtained through $\eta_0 = (dJ(t)/dt)^{-1}$ at terminal flow (inset of **Figure 1e**).[12] After removing the stress at $t = t_0$, a recoverable creep compliance, $J_r(t - t_0) = (\gamma(t_0) - \gamma(t))/\sigma_c$, probes the stored elastic deformation and its relaxation spectrum. $J_r(t - t_0)$ initially follows $J(t)$ through the first compliance plateau $J_1$ upto the characteristic time for the plateau relaxation time $\tau_n$ (**Figure 1e**), where $J_r(t - t_0)$ and $J(t)$ start to deviate. Beyond $\tau_n$, $J_r(t - t_0)$ approaches a second recovery compliance, $J_2$, at long times.

From the two creep compliance, one can estimate the primary strand scale, $M_P$, and a larger molecular scale contributing to long-time recovery, $M_M$, using $M_P \approx \rho RTJ_1$ and $M_M \approx \rho RTJ_2$.[12] At $\sigma_c = 1{,}000$ Pa, PS gives $M_P^{PS} \approx 15\,kg/mol$ and $M_M^{PS} \approx 140\,kg/mol$, whereas P3HT gives $M_P^{P3HT} \approx 10\,kg/mol$ and $M_M^{P3HT} \approx 425\,kg/mol$. For PS, $M_M^{PS}$ remains comparable to $M_w^{PS}$, and within the molecular-weight distribution measured by SEC (**Figure S2b**), consistent with previous observations in polydisperse entangled melts.[28, 29] As shorter chains relax, they progressively dilute the entanglement network formed by the remaining longer chains. In contrast, $M_M^{P3HT} \sim 20 M_w^{P3HT}$, where the SEC trace shows negligible chain population (**Figure 1c**). The long-lived recoverable elasticity therefore requires elastic connectivity extending to multiple chains. Therefore, despite their appearance of similar modulus plateaus, P3HT and PS exhibit fundamentally different network connectivity.

**Plateau relaxation is decoupled from segmental and chain dynamics**

To determine what controls relaxation of the elastic constraints underlying the modulus plateau, we measured creep and creep-recovery over a range of temperatures (**Figures. 2a** and **S6**). For entangled PS, $J(t)$ and $J_r(t - t_0)$ superpose and TTS holds, while $J_1$ remains nearly temperature independent (inset of **Figure S6b**). In contrast, $J(t)$ and $J_r(t - t_0)$ of P3HT change systematically with temperature, consistent with the breakdown of TTS observed in the SAOS. $J_1$ of P3HT also decreases upon heating (inset of **Figure S6a**), corresponding to an increase in plateau modulus observed in **Figure 1d**.

A more fundamental distinction emerges from the temperature dependence of the plateau-relaxation time, $\tau_n$. In PS, as in conventional entangled polymer melts, $\tau_n$ remains coupled to segmental relaxation time, $\tau_\alpha^S \approx a_T^S \tau_\alpha(T_{ref})$, or zero-shear viscosity, $\eta_0$ (**Figure S7**),[11, 30, 31] with super-Arrhenius temperature dependence. In contrast, $\tau_n$ of P3HT follows an Arrhenius temperature dependence (**Figure 2b**). The plateau relaxation is therefore decoupled from both segmental friction and terminal relaxation, indicating a distinct thermally activated process. This decoupling also provides an explanation to the observed TTS breakdown in the plateau regime.

Fitting $\tau_n$ to an Arrhenius relation gives an apparent activation energy $E_a \approx 160\ kJ/mol$ for P3HT. A plausible microscopic origin is the reversible association between neighboring conjugated backbone segments.[32] Interestingly, previous quantum calculations for stacked oligothiophenes report $\pi - \pi$ binding energy of $\sim 30 - 70\ kJ/mol$, depending on oligomer length and stacking geometry.[33, 34] Although these binding energies cannot be compared directly with the apparent activation energy measured rheologically, the substantially larger $E_a$ suggests that network relaxation may involve cooperative rearrangement of multiple interacting thiophene units. Such a picture is also consistent with association over a finite backbone segment of multiple repeating units.[35] These results thus support a transient network whose relaxation is governed by thermally activated reorganization of interchain associations, where $\tau_n$ represents a characteristic time for the reorganization of association-mediated network.

## Stress-dependent recovery reveals an extended interchain network

In polymer melt with topological entanglement, an elastic strand is a subchain confined between neighboring topological constraints and therefore cannot exceed the molecular weight of an individual chain.[11] No such restriction applies to a network formed through interchain association, where reversible junctions can connect different chains into larger elastic structures.[20, 36] Because $J_2$ weighted toward the slowest recoverable modes and is sensitive to long-lived elastic connectivity, creep-recovery measurements provide a probe of this extended network struture.[12, 37, 38]

We therefore measured creep and creep recovery of P3HT at $T = 318\ K$ under $\sigma_C = 100\ Pa$, $500\ Pa$, $1{,}000\ Pa$, $2{,}000\ Pa$, $3{,}000\ Pa$, and $5{,}000\ Pa$ (**Figure 2c**). Analogous measurements were performed on PS at $\sigma_C = 1{,}000\ Pa$ and $5{,}000\ Pa$ at $T = 383\ K$ (**Figure S8**). Nearly identical creep curves are observed at different $\sigma_C$ in a sequence of the measurements for both PS and P3HT, especially at short times, suggesting the absence of creep-induced irreversible structural changes, such as strain-induced crystallization. As expected for the entangled reference, $J(t)$ and $J_r(t - t_0)$ of PS superpose at the two stresses. In contrast, $J(t)$ and $J_r(t - t_0)$ of P3HT remain nearly stress independent only upto $\tau_n$, implying that $J_1$ and $\tau_n$ are only weakly affected by the applied stress. Beyond $\tau_n$, $J(t)$ and $J_r(t - t_0)$ become strongly stress dependent: a lower stress gives a larger $J_2$. At $\sigma_C = 100\ Pa$, $J_2 \sim 8.1 \times 10^{-4}\ Pa^{-1}$ which gives $M_M^{P3HT} \sim 2.37 \times 10^6\ g/mol$ that is nearly $100 M_w^{P3HT}$ (**Figure 2d**). The SEC distribution contains essentially no chains at this molecular-weight scale (**Figure 1c**). The long-time recoverable response therefore requires elastic connectivity extending across many P3HT chains (**Figure 2e**).

At higher stresses, $J_2$ approaches an apparent limiting value with a crossover $\sigma_c \sim 1{,}000 - 3{,}000\ Pa$ (**Figure 2d**). Such stress dependence is not expected for a purely topological entanglement network and instead indicates a distribution of mechanical stability within the interchain network. Weaker junctions are progressively disrupted or reorganized as stress increases, whereas more persistent junctions survive higher stress.

From this perspective, the crossover at $\sigma_c \sim 1{,}000 - 3{,}000\ Pa$ defines a characteristic cohesive stress scale of the interchain network at $318\ K$.

The above revealed presence of interchain network in P3HT raises three possible origins to its modulus plateau in the linear viscoelastic spectra: (i) a topological entanglement dominated network with additional sparse interchain network constraints; (ii) an interchain association-dominated network with sparse topological entanglement; or (iii) a mixed network of topological entanglement and interchain association contributed comparably. The stress-dependent creep recovery measurements demonstrate a substantial interchain contribution but, by itself, do not distinguish between the latter two cases. The primary plateau relaxes through an Arrhenius process and decouples strongly from segmental relaxation and terminal relaxation,[11] highlighting a strong contribution of the interchain association to the network elasticity. Importantly, no second plateau attributable to topological entanglement is resolved between the relaxation of the primary network and terminal flow. It is unlikely that the topological entanglement and the interchain association have comparable contribution to the network plateau elasticity. These analyses therefore favor an interchain-association-dominated network for the modulus plateau in P3HT, although a weaker topological entanglement contribution cannot be excluded.

**Mechanical perturbation reveals persistent reorganization of the plateau network**

Unlike topological entanglements, which arises from chain uncrossability,[7] reversible interchain junctions can break and reform under deformation, allowing the network connectivity to reorganize.[19, 20, 39-41] If the P3HT plateau is governed primarily by such associations, deformation-induced changes in the network should persist after the perturbation is removed and remain visible in the subsequent linear response.

We tested this idea using a sequential SAOS/MAOS (medium amplitude oscillatory shear)/SAOS protocol with stage durations of $t_1$, $t_2$, and $t_3$ (**Figure 3a**). The first SAOS stage establishes the equilibrium linear response, MAOS perturbs the network, and the final SAOS stage probes the post-deformation state.[27] Measurements were performed at $T = 328\ K$ for P3HT and $T = 403\ K$ for PS, where two materials have comparable terminal relaxation times of $\tau_d \sim 10\ s$. The perturbation was applied at $\omega = 10\ s^{-1}$, with medium strain amplitudes of $\gamma_0 = 10\%$ for P3HT and $\gamma_0 = 20\%$ for PS, chosen near the end of the linear response region (**Figure S9**). A partitioned-plate (PP) geometry was used to minimize edge fracture (see **MATERIALS AND METHODS**).

PS shows little change during the $t_2 = 1200\ s \approx 120\ \tau_d$ perturbation, and both $G'$ nor $G''$ return rapidly to their initial values when the experiment returns to SAOS (**Figure S10**). This is expected as $\gamma_0$ at the MAOS region is in the linear response region. In contrast, $G'$ of P3HT *increases* first from approximately 0.43 to 0.47 MPa over the first $200\ s$, and then *decreases* progressively (upto 13%) with continued deformation during MAOS (**Figure 3b**). The initial stiffening followed by softening indicates substantial evolution of the network under sustained deformation. $G''$ and $tan(\delta)$ evolve concurrently (**Figure 3b**), further indicating a progressive change in the viscoelastic state.

The distinction becomes even clearer after the MAOS perturbation is removed. Both $G'$ and $G''$ of P3HT remain significantly below their equilibrium values and do not recover even after $\sim 10^2 \tau_d$. The MAOS step therefore leaves P3HT in a persistent post-deformation state rather than producing only a transient nonlinear distortion. Full recovery of $G'$ and $G''$ occurs after a much longer time, $\sim 10^3 \tau_d$, and at a higher testing temperature (**Figure S11**). Thus, deformation drives P3HT into a long-lived metastable viscoelastic state whose recovery is much slower than the terminal chain relaxation. These observations indicate the presence of long-lived elastic constraints in conjugated polymers, which are consistent with the long-time elastic recovery, $J_2$, associated with the multi-chain connectivity (**Figure 2e**).

Frequency sweeps after MAOS provide a more complete view of this reorganized state (**Figures 3c** and **3d**). Both $G'$ and $G''$ remain reduced, while the modulus plateau extends over a broader frequency range than in the equilibrium spectrum. The lower plateau modulus indicates fewer elastically active constraints, whereas extension of the plateau to lower frequencies indicates a longer-lived network. The accompanying change in $tan\delta$ further supports a genuine change in the material response rather than a mechanical artifact such as edge fracture.

The post-deformation state therefore exhibits a lower modulus along with a longer modulus plateau. The combination is difficult to reconcile with a purely topological-entanglement picture, where topological entanglement has a fast recovery dynamics after distortion (**Figure 4a**). On the other hand, these observations follow naturally from reorganization of a transient interchain network: Deformation can remove or rearrange weaker junctions, reducing the number of elastically active constraints, while leaving behind a more persistent population of connections (**Figure 4b**). The resulting network has a larger effective spacing between active junctions and a longer lifetime of the interchain association constraints.

The long memory of the deformation-induced state of P3HT, combined with the Arrhenius relaxation of $\tau_n$ and the extended interchain connectivity revealed by creep recovery, supports an interchain association-dominated network origin of the P3HT plateau, which is consistent with the reversible $\pi - \pi$ interactions between conjugated backbone segments.

**Side-chain length tunes plateau elasticity and relaxation in P3ATs**

To test whether the behavior observed in P3HT extends across the P3AT family, we performed creep and creep-recovery measurements on two additional amorphous regiorandom P3ATs, poly(3-butylthiophene-2,5-diyl) (P3BT) and poly(3-octylthiophene-2,5-diyl) (P3OT). These polymers share the same conjugated backbone as P3HT but differ in the alkyl side-chain length from P3HT.

Like P3HT, P3BT and P3OT exhibit a well-defined modulus plateau and a breakdown of TTS in the plateau regime (**Figures 5a** and **5b**, **Figure S12**). The plateau compliance, $J_1$, increases systematically with side-chain length with an order $J_{1,P3BT} < J_{1,P3HT} < J_{1,P3OT}$ (**Figure S13**). Similar side-chain-dependent changes in modulus have also been observed in conventional polymers,[5, 42] so this trend alone does not distinguish associative from topological constraints. It nevertheless shows that the density of elastic constraints in P3ATs is strongly coupled to side-chain structure.

A more distinctive trend appears in the plateau-relaxation dynamics. For both P3BT and P3OT, $\tau_n$ follows an Arrhenius temperature dependence, whereas the segmental and zero-shear viscosity, $\tau_\alpha^S$ and $\eta_0$, remain super-Arrhenius (**Figures 5c** and **5d**). The decoupled, thermally activated relaxation observed in P3HT therefore persists across the P3AT series.

Furthermore, the apparent activation energy associated with $\tau_n$ also decreases systematically with increasing side-chain length, from $E_a \approx 190\ kJ\ mol^{-1}$ of P3BT, $E_a \approx 160\ kJ\ mol^{-1}$ of P3HT, to $E_a \approx 135\ kJ\ mol^{-1}$ of P3OT. Therefore, side-chain length alters not only the magnitude of the plateau elasticity but also the activated process governing its relaxation. Longer alkyl side chains are expected to reduce close contact between conjugated backbones through increased steric separation. The concurrent decrease in plateau modulus and activation barrier is thus consistent with progressively weaker or less extensive backbone association. These results extend the physical picture established for P3HT across P3AT series: the plateau arises from elastic constraints whose relaxation is thermally activated and decoupled from segmental and terminal dynamics, while both the number density and lifetime of these interchain constraints are tunable through side-chain structure.

## DISCUSSION

Our results support a reversible interchain network as the dominant origin of the modulus plateau in these P3ATs, although a contribution from topological entanglement cannot be excluded. The key distinction from a conventional entangled melt is not the presence of a plateau itself, but in the dynamics and connectivity of the constraints that sustain it. In P3ATs, the constraints extend across multiple chains, relax through a thermally activated process decoupled from segmental and terminal dynamics, and can be reorganized by external deformation. These behaviors are consistent with long-lived interchain associations between conjugated backbone segments.

This distinction has an important consequence for the interpretation of plateau rheology. The plateau modulus, $G_N$, reflects the density of long-lived elastic constraints but does not by itself identify their microscopic origin. When reversible associations contribute substantially to $G_N$, the molecular scale inferred from $G_N$ represents an effective network-constraint scale rather than a purely topological entanglement molecular weight, $M_e$.[18, 20-22] The same caution applies to plateau-width scaling and related analyses that interpret the plateau within a purely topological framework.[11, 12, 43] P3ATs thus provide a clear counterexample to treating the modulus plateau as an unambiguous signature of topological entanglement, a caution that is particularly important because conjugated polymers—with reported backbone persistence lengths of approximately 2–30 nm[44-46]—have been actively used as model systems for semiflexible entanglement.[16, 17] The same molecular architecture, however, also enables interchain associations capable of generating long-lived elastic constraints.[47, 48] Associative contributions must therefore be considered before plateau rheology is interpreted solely in terms of topological entanglement.

The present measurements also illustrate how these contributions can be distinguished experimentally. Linear rheology establishes the presence of long-lived elasticity, while creep recovery probes the molecular scale over which elastic connectivity persists. Temperature-dependent relaxation tests whether the plateau follows segmental friction or a distinct activated process, and mechanical perturbation probes whether the network can be reorganized independently of terminal relaxation. No single measurement uniquely identifies the microscopic origin of the plateau, but their combination places substantially stronger constraints on its interpretation far more strongly than $G_N$ alone.

The association-mediated elasticity also introduces a molecular variable absent from a purely topological description. Across P3AT series, side-chain length changes both the plateau modulus and the activated relaxation of the plateau-forming network, showing that molecular structure tunes not only the density of elastic constraints but also their dynamics. Because $\pi - \pi$ interactions are already known to influence backbone packing, interchain electronic coupling, and charge transport in conjugated polymers,[49-52] their contribution to mechanical relaxation provides an additional connection between molecular structure and processing behavior. The strength and lifetime of interchain association therefore emerge as useful design parameters for controlling the viscoelastic response of conjugated polymers.

## MATERIALS AND METHODS

### Materials

Regiorandom poly(3-butylthiophene-2,5-diyl) (P3BT), regiorandom poly(3-hexylthiophene-2,5-diyl) (P3HT), and regiorandom poly(3-octylthiophene-2,5-diyl) (P3OT) were purchased from Rieke Metals. The average molecular weight (MW) and polydispersity (*Đ*) were characterized by size exclusion chromatography. P3BT has a weight-averaged MW, $M_w^{P3BT}$, of $37\ kg/mol$ and *Đ* = 2.0. P3HT has $M_w^{P3HT}$ of $21\ kg/mol$ and *Đ* = 2.4. P3OT has $M_w^{P3OT}$ of $27\ kg/mol$ and *Đ* = 2.3. Polystyrene (PS) was obtained from Dow Chemical, with $M_w^{PS}$ of $172\ kg/mol$ and *Đ* = 2.2. All materials were used after drying in a vacuum oven for 24 hours before measurements without further purification.

### Small- and wide-angle x-ray scattering (SAXS/WAXS)

Small- and wide-angle x-ray scattering (SAXS/WAXS) measurements were performed at the beamline 12-ID-B of Advanced Photon Source at Argonne National Laboratory. An x-ray with a wavelength of 0.9322 Å was used. SAXS patterns were recorded with an in-vacuum Eiger2 9M (DECTRIS) detector with an array of 0.075 mm square pixels in a 3262 × 3108 configuration and the distance between the sample and detector was kept at 2.0 m. WAXS patterns were recorded using an in-vacuum Pilatus300K detector (DECTRIS) with an array of 0.172 mm square pixels in a 619 × 487 configuration and the distance between the sample and detector was kept at 0.42 m. The scattering vector $q$ was calibrated with silver behenate, and the absolute scattering intensity was calibrated using glassy carbon. The materials were sealed into a DSC aluminum pan. An empty DSC aluminum pan was measured for reference to eliminate the signal influence from the aluminum pan. The x-ray measurements were performed from room temperature to 533 K at an interval of 20 K. Azimuthal integration was applied to convert the 2D scattering images to 1D intensity profiles using the beamline software.

**Size exclusion chromatography (SEC)**

Size exclusion chromatography (SEC) was performed on Agilent PL-GPC. Chloroform was used as the eluent, and the measurement temperature was kept at 313 K. The SEC was calibrated with polystyrene (PS) standards with MW ranging from 162 to 364,000 g/mol with chloroform as the eluent at 313 K (**Figure S1b**), from which a universal calibration curve is further constructed. Specifically, $[\eta] = KM^{\alpha+1}$, where $K_{PS} = 7.15 \times 10^{-5}\ dL/g$ and $\alpha_{PS} = 0.76$ for polystyrene in chloroform.[53] The molecular weight of P3HT with an elution time $t_i$, $M_{P3HT,t_i}$, is calculated through the universal calibration curve, $M_{P3HT,t_i} = \left(\frac{K_{PS} M_{PS,t_i}^{\alpha_{PS}+1}}{K_{P3HT}}\right)^{\frac{1}{\alpha_{P3HT}+1}}$, where $M_{PS,t_i}$ is the molecular weight of the standard PS at the elution time, $t_i$. For P3HT in chloroform, $K_{P3HT} = 3.32 \times 10^{-5}\ dL/g$ and $\alpha_{P3HT} = 0.955$.[54] Due to the limited data for P3BT and P3OT in literature and their similar properties as P3HT, we use the literature value for P3HT in chloroform to approximately analyze the SEC traces of P3BT and P3OT. The weight fraction distribution of the testing sample can be further calculated as $w_i = \frac{RI_i}{\sum_i RI_i}$, where $w_i$ stands for weight fraction of the corresponding polymer at the elution time $t_i$, and $RI_i$ is its corresponding intensity (**Figure 1b** and **Figure S2**).

**Differential scanning calorimetry (DSC)**

The thermal properties of the samples were characterized using differential scanning calorimetry (DSC Q2000, TA Instruments). A heat-cool cycle was performed over a prescribed temperature range at a constant heating and cooling rate of 10 °C/min. Three cool-heat cycles were performed to check the reproducibility

**Small-amplitude oscillatory shear (SAOS)**

Small-amplitude oscillatory shear (SAOS) measurements were performed at Anton Paar MCR 302. The rheometer is equipped with a Peltier temperature controller providing a temperature accuracy of $\pm 0.05\,K$. The sample was sandwiched between a pair of aluminum parallel plates with a diameter of 4 mm and with a fixed gap of 1.0 mm. A strain amplitude of 0.1% was applied to probe the linear viscoelastic response near $Tg$. The strain amplitude was gradually increased to 5% to access lower moduli (below $10^6\ Pa$) in the terminal flow regime.

**Creep and creep recovery rheology.**

Creep and creep recovery measurements were performed using an Anton Paar MCR 302 rheometer equipped with a Peltier temperature control system with a temperature accuracy of $\pm 0.05\,K$. The sample was sandwiched between two parallel aluminum plates with a diameter of 4 mm and a fixed gap of 1.0 mm. A constant shear stress, $\sigma$, was applied for a preset duration, $t_1 = 8.64 \times 10^5\ s$, after which the stress was removed, allowing the sample to recover freely for a preset duration, $t_2 = 8.64 \times 10^5\ s$. The resulting strain response, $\gamma$, was continuously recorded throughout both the creep and recovery stages.

**SAOS/MAOS/SAOS**

Strain-controlled SAOS/MAOS/SAOS measurements were performed using an Advanced Rheometric Expansion System (ARES, TA Instruments) equipped with a Rheometric Scientific Oven for temperature control with an accuracy of $\pm 0.05\,K$. Measurements were conducted using partitioned-plate geometry, with the sample sandwiched between a top partitioned plate and a bottom parallel plate at a gap of 1.0 mm. The top partitioned plate had an inner measuring diameter of 6 mm and an outer ring diameter of 15 mm, while a 25 mm in diameter parallel plate was used as the bottom plate. A sequential three-stage time-sweep oscillatory shear protocol was performed at a constant angular velocity $\omega$, with the strain amplitude $\gamma_0$ and duration $t$ prescribed independently for each stage.

**Supplementary Materials**

**The file includes:**

Figures S1-S2: SEC characterization and analysis;

Figure S3: DSC for P3ATs;

Figure S4: Temperature-dependent SAXS/WAXS of P3BT and P3OT;

Figure S5: Dynamic shift factors and modulus shift factors of P3HT and PS;

Figure S6: Temperature-dependent creep and creep recovery of P3HT and PS;

Figure S7: Temperature dependence of characteristic relaxation times in PS;

Figure S8: Creep and creep recovery of PS under varying stress;

Figure S9: Strain sweep of P3HT at $T = 328\ K$ and PS at $T = 403\ K$ under $\omega = 10\ s^{-1}$;

Figure S10: Sequential SAOS/MAOS/SAOS measurement on PS at $T = 403\ K$ ;

Figure S11: Sequential SAOS/MAOS/SAOS measurement on P3HT at $T = 373\ K$ ;

Figure S12: Master curves, Dynamic shift factors, and van-Gurp plots of P3BT and P3OT;

Figure S13: Temperature-dependent creep response of P3BT and P3OT.

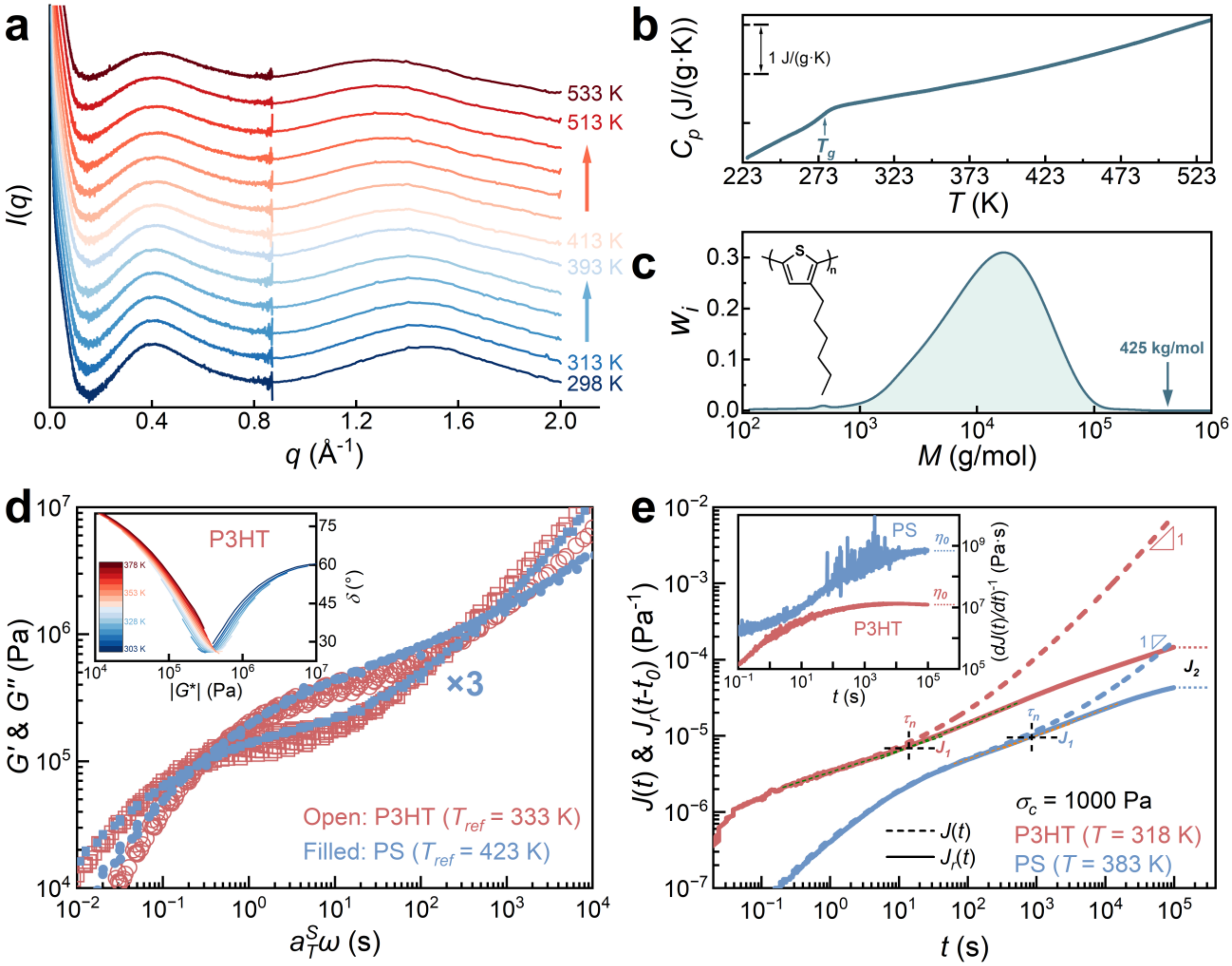

**Figure1. Structural and viscoelastic characterizations on P3HT. (a)** Small-angle x-ray scattering and wide-angle x-ray scattering (SAXS/WAXS) of P3HT at the temperature of $298\ K - 533\ K$ at an interval of $20\ K$. No crystalline or liquid crystalline ordering has been observed during the measurements. **(b)** Specific heat capacity, $C_p$, of P3HT upon second heating with a ramping rate of $10\ K/min$ at the temperature of $223\ K - 533\ K$. **(c)** Molecular weight distribution of P3HT from the size exclusion chromatography (SEC). **(d)** Linear viscoelasticity master curves of P3HT (red open symbols) at a reference temperature of $T_{ref} = 333\ K$ and PS (blue filled symbols) at $T_{ref} = 423\ K$. The dynamic moduli of PS are shifted upward by a factor of 3 for comparison. The inset shows the corresponding van Gurp-Palmen plot for P3HT, the phase angle $\delta$ versus absolute complex modulus $|G^*|$, across the measured temperature range, where clear breakdown of the time-temperature superposition can be observed at the modulus range of $10^5 - 10^6$ Pa. **(e)** Creep compliance, $J(t)$ (dashed lines), and recoverable creep compliance, $J_r(t - t_0)$ (solid lines), for P3HT (red) at $318\ K$ and PS (blue) at $383\ K$ under a creep stress of $\sigma_C = 1000\ Pa$. The compliance plateau is observed with value $J_1$ that is identified at $\tau_n$, where $J(t)$ and $J_r(t - t_0)$ deviate. Creep recovery completes at a higher-compliance plateau, $J_2$. The inset shows the corresponding time-dependent $(dJ(t)/dt)^{-1}$ upon creep and the determination of $\eta_0$.

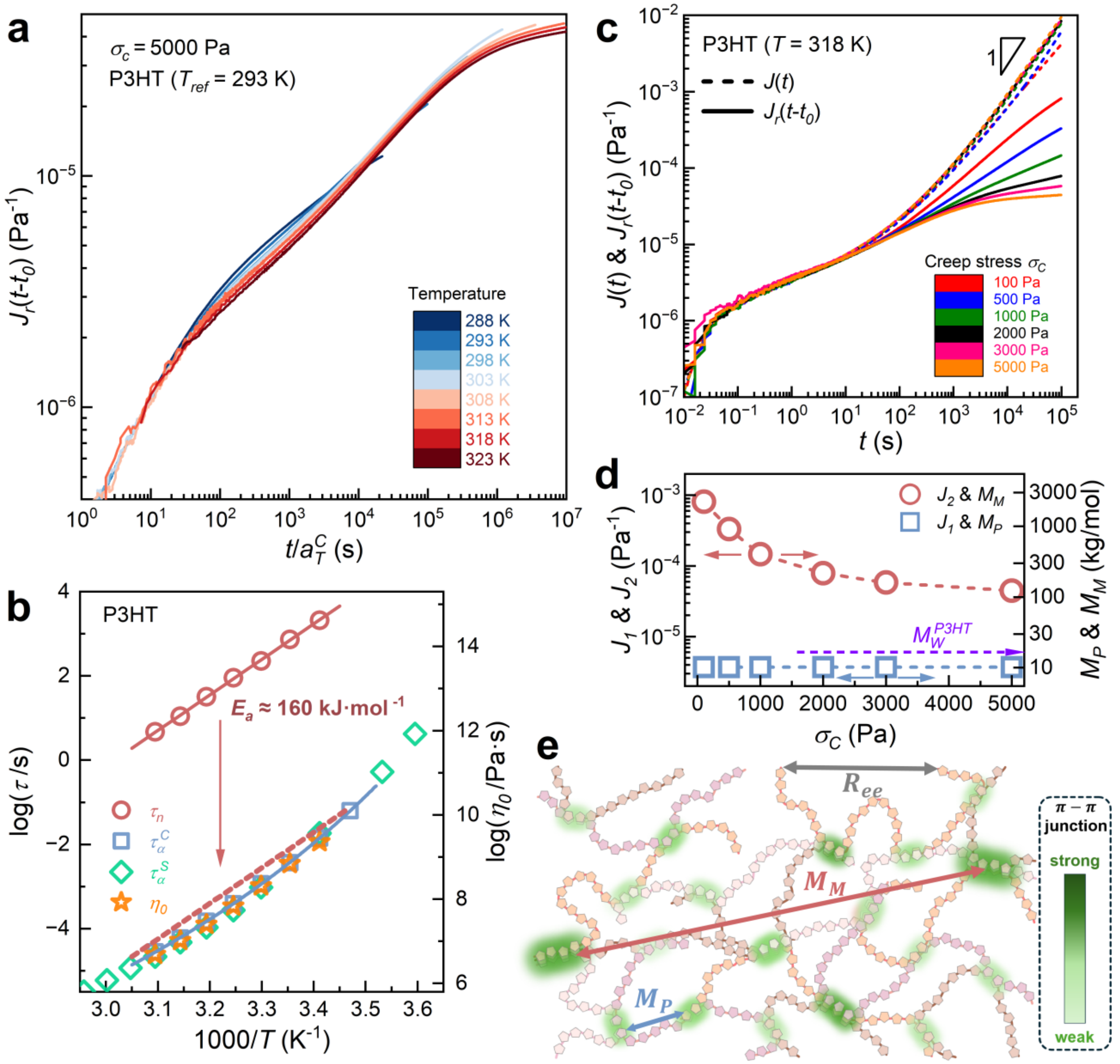

**Figure 2. Plateau elasticity and hierarchical network mediated by interchain $\pi-\pi$ interactions. (a)** Recoverable creep compliance, $J_r(t-t_0)$, of P3HT following a creep at $\sigma_C = 5000\,Pa$ over the temperature range of $288\,K - 323\,K$ with $T_{ref} = 293\,K$. **(b)** Temperature dependences of plateau-relaxation time (red circles), $\tau_n$, segmental relaxation times derived from creep measurement (blue squares), $\tau_\alpha^C$, and from SAOS (green diamonds), $\tau_\alpha^S$, and zero-shear viscosity (orange stars), $\eta_0$. The red solid line presents the Arrhenius fit to $\tau_n$, and the blue solid line presents the VFT fit to $\tau_\alpha^C$. The red dashed line shows the Arrhenius fit shifted vertically to highlight the differences in temperature dependences of $\tau_n$ and $\tau_\alpha^C$. **(c)** Creep compliance (dashed lines), $J(t)$, and recoverable creep compliance (solid lines), $J_r(t-t_0)$, of P3HT at $318\,K$ under different

creep stresses, $\sigma_C$. **(d)** $J_1$ (blue squares) and $J_2$ (red circles) and the molecular weights of their corresponding strands, $M_P$ and $M_M$ at different $\sigma_C$. The purple arrow indicates the average molecular weight of P3HT ($M_W^{P3HT}$) from SEC. **(e)** A schematic illustration of the network connectivity of conjugated polymers that leads to $J_1$ and $J_2$. Reversible interchain $\pi - \pi$ junctions with different interaction strength serve as network constraints. The average spacing between neighboring interchain $\pi - \pi$ junctions defines the primary network stand length, $M_P$ (blue arrow). The larger network with size $M_M$ (red arrow) connecting multiple polymer chains through reversible interchain $\pi - \pi$ junctions gives the characteristic long-time recovery, $J_2$. $R_{ee}$ denotes the end-to-end distance (grey arrow) of an individual chain. Color bar shows the differences in the strength of $\pi - \pi$ junctions. Different colors of chains are used to distinguish different chains in a conjugated homopolymer melt.

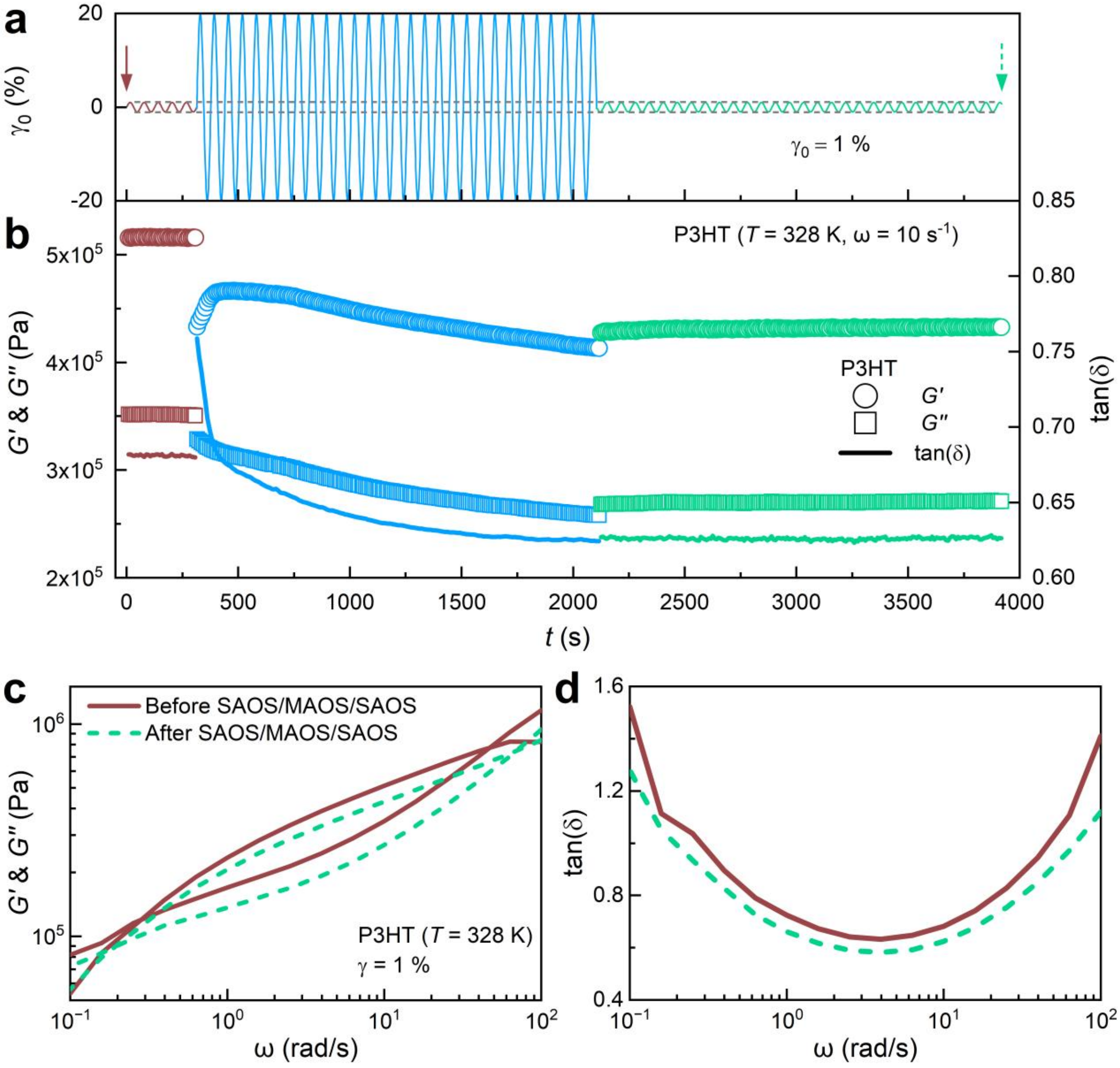

**Figure 3. Persistent reorganization of plateau network in P3HT under strain. (a)** A schematic illustration of SAOS/MAOS/SAOS measurement on P3HT at $328\,K$ at a constant angular frequency $\omega = 10\ s^{-1}$. A strain amplitude of $\gamma_0 = 1\%$ is applied to the initial (red) and the final (green) SAOS stages and a $\gamma_0 = 10\%$ is employed in the middle MAOS (blue) stage. **(b)** Time-dependent dynamic moduli, $G'$ (circles) and $G''$ (squares), and the loss factor, $tan(\delta)$ (solid line), of P3HT upon SAOS/MAOS/SAOS measurements. **(c)** Dynamic moduli, $G'$ and $G''$, and **(d)** loss factor, $tan(\delta)$, of a frequency sweep SAOS at $\gamma = 1\%$ on P3HT before (red solid lines) and after (green dashed lines) the SAOS/MAOS/SAOS measurement. After the MAOS perturbation, a clear reduction in plateau modulus along with a lengthening in the plateau width has been observed.

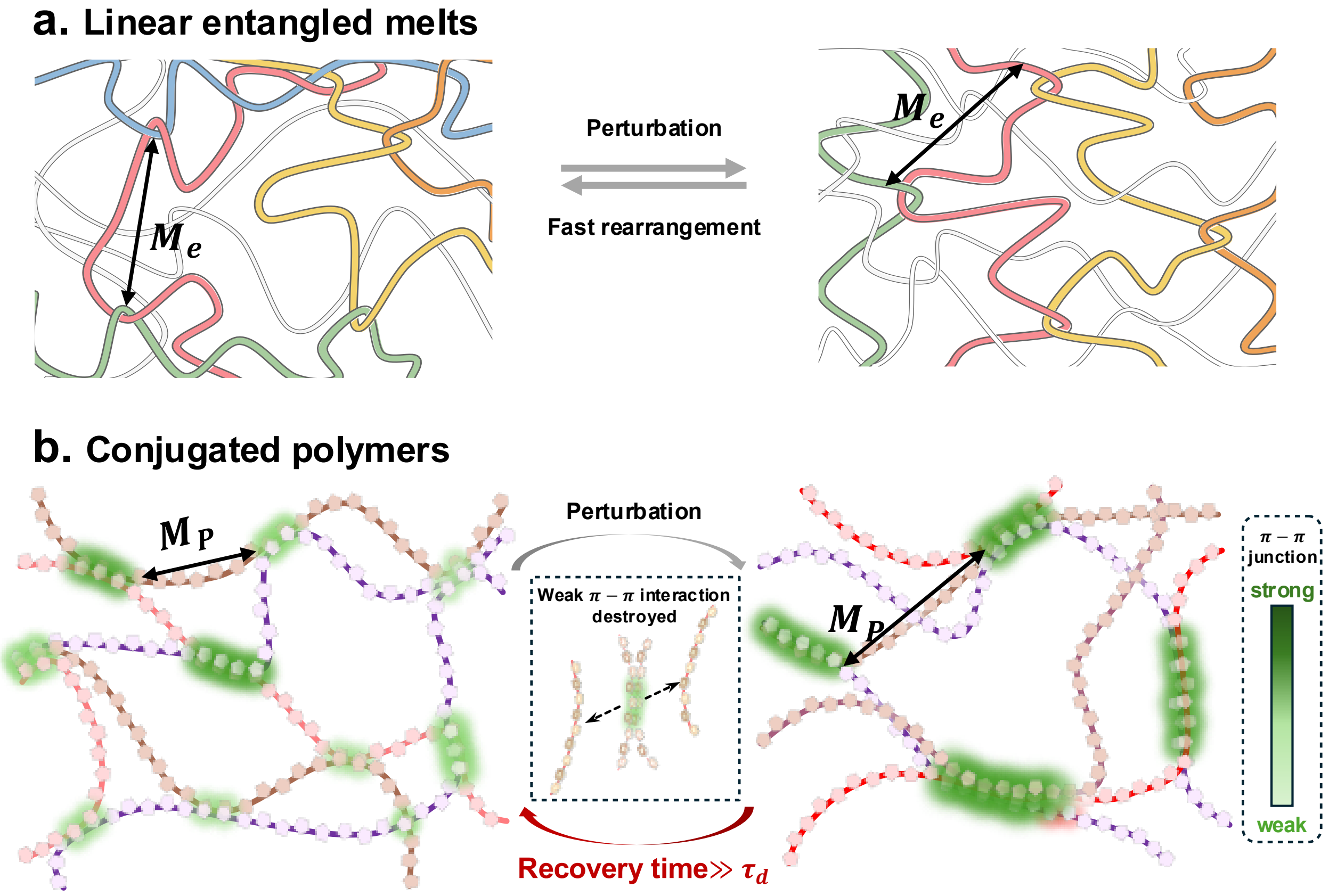

**Figure 4. Mechanism of network reorganization upon mechanical perturbation. (a)** An illustration of structural changes of linear entangled polymers to external perturbation. Entanglements may be distorted during deformation. A fast rearrangement and recovery is observed for the entanglement molecular weight $M_e$, upon removal of the external perturbation. **(b)** An illustration of structural changes of conjugated polymers to external perturbation. $\pi-\pi$ interaction-mediated network initially has a broad distribution of $\pi-\pi$ junction strength (left), represented by the brightness strength and sizes variation in the green-shaded region. Upon external deformation, weak $\pi-\pi$ junctions (light colored ones) can be destructed (middle) or rearranged. At the same time, the deformation-induced chain reorganization can induce formation of $\pi-\pi$ junctions with stronger interaction strength (right panel). The loss of the weak $\pi-\pi$ junctions increases $M_P$, and reduces the plateau modulus. The formation of stronger $\pi-\pi$ junctions gives rise to an lengthening in plateau width. The time required for recovery to initial state is significantly longer than terminal relaxation time, $\tau_d$, in conjugated polymers. Different colors of chains are used only to distinguish different polymer chains in a conjugated homopolymer melt. Color bar shows the strength distribution of $\pi-\pi$ junctions.

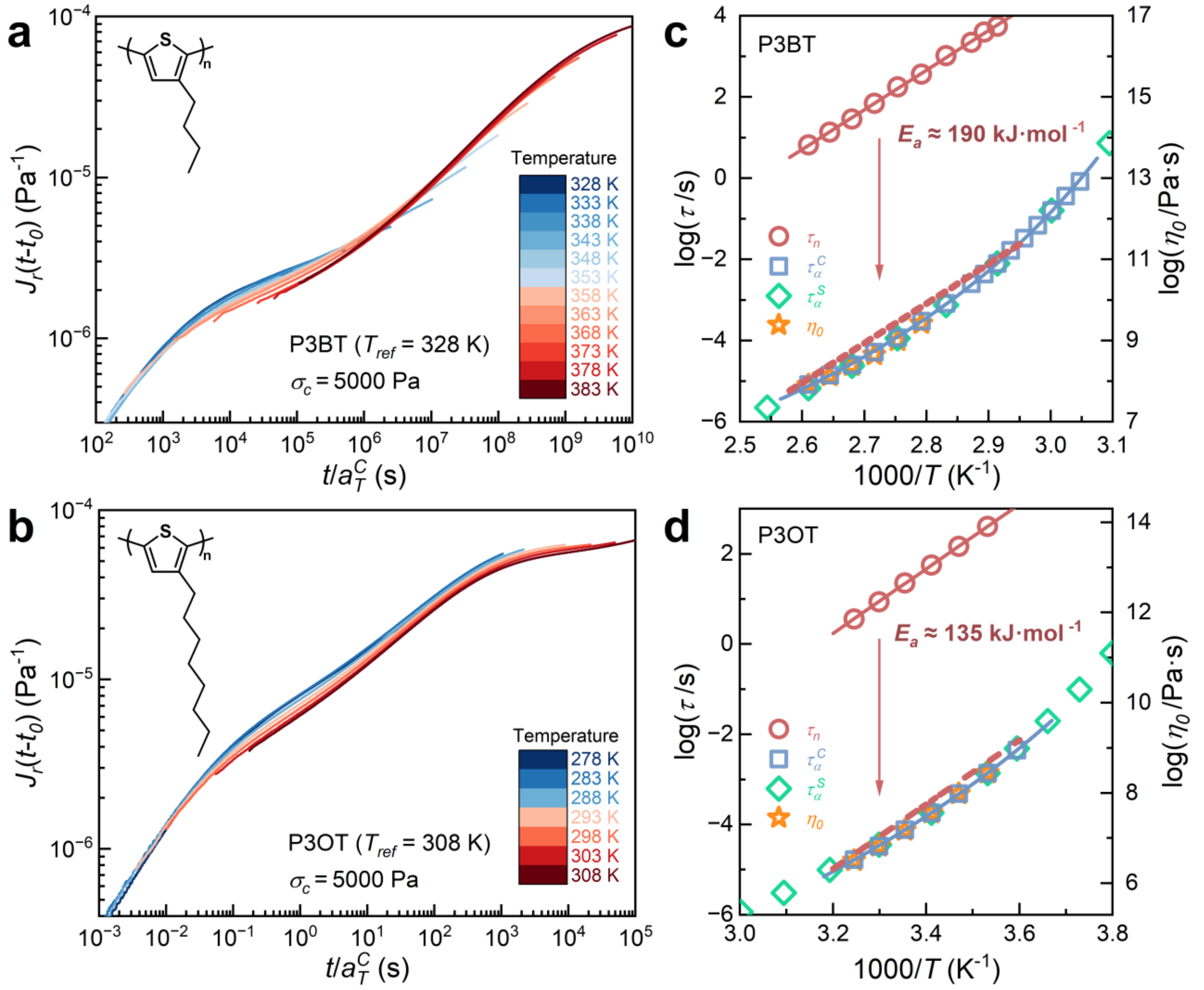

**Figure 5. Universality of $\pi-\pi$ interaction-mediated networks in P3ATs. (a-b)** Recoverable creep compliance, $J_r(t-t_0)$, after a creep test at a stress of $\sigma_C = 5000\ Pa$ for **(a)** P3BT at $328\ K - 383\ K$ with $T_{ref} = 328\ K$ and **(b)** P3OT at $278\ K - 308\ K$ with $T_{ref} = 308\ K$. **(c-d)** Temperature dependences of $\tau_n$ (red circles), segmental relaxation times derived from creep measurement (blue squares), $\tau_\alpha^C$, segmental relaxation time from SAOS (green diamonds), $\tau_\alpha^S$, and zero-shear viscosity (orange stars), $\eta_0$, of **(c)** P3BT and **(d)** P3OT. The red solid lines present the Arrhenius fit to $\tau_n$, and the blue solid lines present the VFT fit to $\tau_\alpha^C$. The red dashed lines show the Arrhenius fit shifted vertically to facilitate comparison and highlight the differences in temperature dependences of plateau relaxation and segmental relaxation.

## References

1. Matsuhisa, N.; Chen, X.; Bao, Z.; Someya, T. Materials and structural designs of stretchable conductors. *Chem. Soc. Rev.* **2019,** 48 (11), 2946-2966 DOI: 10.1039/c8cs00814k.
2. Someya, T.; Bao, Z.; Malliaras, G. G. The rise of plastic bioelectronics. *Nature* **2016,** 540 (7633), 379-385 DOI: 10.1038/nature21004.
3. Zhao, C.; Park, J.; Root, S. E.; Bao, Z. Skin-inspired soft bioelectronic materials, devices and systems. *Nature Reviews Bioengineering* **2024,** 2 (8), 671-690 DOI: 10.1038/s44222-024-00194-1.
4. Wang, Y.; Cao, Z.; Gu, X., Stretchable Conjugated Polymers and Composites for Wearable Electronics. In *Springer Handbook of Functional Polymers*, Chujo, Y., Ed. Springer Nature Singapore: Singapore, 2025; pp 681-707.
5. Fetters, L. J.; Lohse, D. J.; Colby, R. H., Chain Dimensions and Entanglement Spacings. In *Physical Properties of Polymers Handbook*, Mark, J., Ed. Springer New York: 2007; pp 447-454.
6. Hoy, R. S.; Kröger, M. Unified Analytic Expressions for the Entanglement Length, Tube Diameter, and Plateau Modulus of Polymer Melts. *Phys. Rev. Lett.* **2020,** 124 (14), 147801 DOI: 10.1103/PhysRevLett.124.147801.
7. Qin, J.; Milner, S. T. Tubes, Topology, and Polymer Entanglement. *Macromolecules* **2014,** 47 (17), 6077-6085 DOI: 10.1021/ma500755p.
8. Milner, S. T. Unified Entanglement Scaling for Flexible, Semiflexible, and Stiff Polymer Melts and Solutions. *Macromolecules* **2020,** 53 (4), 1314-1325 DOI: 10.1021/acs.macromol.9b02684.
9. Qin, J. Similarity of Polymer Packing in Melts Is Dictated by $\bar{N}$. *Macromolecules* **2024,** 57 (5), 1885-1892 DOI: 10.1021/acs.macromol.3c02273.
10. Graessley, W. The entanglement concept in polymer rheology. *Adv. Polym. Sci.* **1974,** 16, 1-179 DOI: 10.1007/BFb0031037.
11. Doi, M., and S.F. Edwards, *The theory of polymer dynamics*. Oxford University Press, New York: 1986.
12. Ferry, J. D., *Viscoelastic properties of polymers*. Wiley, New York: 1980.
13. Yan, Z.-C.; Li, Y.; Guo, Z.; Shinohara, A.; Nakanishi, T.; Chen, G.; Pan, C.; Stadler, F. J. Rheology of Conjugated Polymers with Bulky and Flexible Side Chains. *Macromolecules* **2021,** 54 (9), 4061-4069 DOI: 10.1021/acs.macromol.1c00044.
14. Xie, R.; Aplan, M. P.; Caggiano, N. J.; Weisen, A. R.; Su, T.; Müller, C.; Segad, M.; Colby, R. H.; Gomez, E. D. Local Chain Alignment via Nematic Ordering Reduces Chain Entanglement in Conjugated Polymers. *Macromolecules* **2018,** 51 (24), 10271-10284 DOI: 10.1021/acs.macromol.8b01840.
15. Xie, R.; Lee, Y.; Aplan, M. P.; Caggiano, N. J.; Müller, C.; Colby, R. H.; Gomez, E. D. Glass Transition Temperature of Conjugated Polymers by Oscillatory Shear Rheometry. *Macromolecules* **2017,** 50 (13), 5146-5154 DOI: 10.1021/acs.macromol.7b00712.
16. Fenton, A. M.; Xie, R.; Aplan, M. P.; Lee, Y.; Gill, M. G.; Fair, R.; Kempe, F.; Sommer, M.; Snyder, C. R.; Gomez, E. D.; Colby, R. H. Predicting the Plateau Modulus from Molecular Parameters of Conjugated Polymers. *ACS Central Science* **2022,** 8 (2), 268-274 DOI: 10.1021/acscentsci.1c01396.

17. Dietz, J. D.; Kröger, M.; Hoy, R. S. Validation and Refinement of Unified Analytic Model for Flexible and Semiflexible Polymer Melt Entanglement. *Macromolecules* **2022,** 55 (9), 3613-3626 DOI: 10.1021/acs.macromol.1c02597.
18. Shabbir, A.; Javakhishvili, I.; Cerveny, S.; Hvilsted, S.; Skov, A. L.; Hassager, O.; Alvarez, N. J. Linear Viscoelastic and Dielectric Relaxation Response of Unentangled UPy-Based Supramolecular Networks. *Macromolecules* **2016,** 49 (10), 3899-3910 DOI: 10.1021/acs.macromol.6b00122.
19. Leibler, L.; Rubinstein, M.; Colby, R. H. Dynamics of reversible networks. *Macromolecules* **1991,** 24 (16), 4701-4707 DOI: 10.1021/ma00016a034.
20. Zhang, Z.; Chen, Q.; Colby, R. H. Dynamics of associative polymers. *Soft Matter* **2018,** 14 (16), 2961-2977 DOI: 10.1039/C8SM00044A.
21. Gold, B. J.; Hövelmann, C. H.; Lühmann, N.; Pyckhout-Hintzen, W.; Wischnewski, A.; Richter, D. The microscopic origin of the rheology in supramolecular entangled polymer networks. *J. Rheol.* **2017,** 61 (6), 1211-1226 DOI: 10.1122/1.4998159.
22. Kim, M.; Patil, S.; Wiegart, L.; Cheng, S.; Cai, L.-H. Associative polymers with controlled sticker placement: How reversible bond distribution and density govern polymer dynamics. *Science Advances* **2026,** 12 (26), eaec4474 DOI: 10.1126/sciadv.aec4474.
23. Shen, X.; Duzhko, V. V.; Russell, T. P. A Study on the Correlation Between Structure and Hole Transport in Semi-Crystalline Regioregular P3HT. *Advanced Energy Materials* **2013,** 3 (2), 263-270 DOI: https://doi.org/10.1002/aenm.201200509.
24. Hugger, S.; Thomann, R.; Heinzel, T.; Thurn-Albrecht, T. Semicrystalline morphology in thin films of poly(3-hexylthiophene). *Colloid. Polym. Sci.* **2004,** 282 (8), 932-938 DOI: 10.1007/s00396-004-1100-9.
25. Xu, Z.; Sun, R.; Lu, W.; Patil, S.; Mays, J.; Schweizer, K. S.; Cheng, S. Nature of Steady-State Fast Flow in Entangled Polymer Melts: Chain Stretching, Shear Thinning, and Viscosity Scaling. *Macromolecules* **2022,** 55 (23), 10737-10750 DOI: 10.1021/acs.macromol.2c01345.
26. Costanzo, S.; Huang, Q.; Ianniruberto, G.; Marrucci, G.; Hassager, O.; Vlassopoulos, D. Shear and Extensional Rheology of Polystyrene Melts and Solutions with the Same Number of Entanglements. *Macromolecules* **2016,** 49 (10), 3925-3935 DOI: 10.1021/acs.macromol.6b00409.
27. Yang, J.; Melton, M.; Sun, R.; Yang, W.; Cheng, S. Decoupling the Polymer Dynamics and the Nanoparticle Network Dynamics of Polymer Nanocomposites through Dielectric Spectroscopy and Rheology. *Macromolecules* **2020,** 53 (1), 302-311 DOI: 10.1021/acs.macromol.9b01584.
28. Plazek, D. J. The Temperature Dependence of the Viscoelastic Behavior of Poly(vinyl acetate). *Polym. J.* **1980,** 12 (1), 43-53 DOI: 10.1295/polymj.12.43.
29. Link, G.; Schwarzl, F. R. Shear creep and recovery of a technical polystyrene. *Rheol. Acta* **1987,** 26 (4), 375-384 DOI: 10.1007/BF01332257.
30. Gennes, P.-G. d., *Scaling concepts in polymer physics*. Cornell University Press, Ithaca and London: 1979.
31. Larson, R. G.; Sridhar, T.; Leal, L. G.; McKinley, G. H.; Likhtman, A. E.; McLeish, T. C. B. Definitions of entanglement spacing and time constants in the tube model. *J. Rheol.* **2003,** 47 (3), 809-818 DOI: 10.1122/1.1567750.

32. Shen, X.; Hu, W.; Russell, T. P. Measuring the Degree of Crystallinity in Semicrystalline Regioregular Poly(3-hexylthiophene). *Macromolecules* **2016,** 49 (12), 4501-4509 DOI: 10.1021/acs.macromol.6b00799.
33. Rodríguez-Ropero, F.; Casanovas, J.; Alemán, C. Ab initio calculations on π-stacked thiophene dimer, trimer, and tetramer: Structure, interaction energy, cooperative effects, and intermolecular electronic parameters. *J. Comput. Chem.* **2008,** 29 (1), 69-78 DOI: https://doi.org/10.1002/jcc.20763.
34. Pickholz, M.; dos Santos, M. C. Theoretical investigation on π-dimer formation in oligothiophenes. *J. Mol. Struct. THEOCHEM* **2005,** 717 (1), 99-106 DOI: https://doi.org/10.1016/j.theochem.2004.11.001.
35. McCulloch, B.; Ho, V.; Hoarfrost, M.; Stanley, C.; Do, C.; Heller, W. T.; Segalman, R. A. Polymer Chain Shape of Poly(3-alkylthiophenes) in Solution Using Small-Angle Neutron Scattering. *Macromolecules* **2013,** 46 (5), 1899-1907 DOI: 10.1021/ma302463d.
36. Rubinstein, M.; Semenov, A. N. Thermoreversible Gelation in Solutions of Associating Polymers. 2. Linear Dynamics. *Macromolecules* **1998,** 31 (4), 1386-1397 DOI: 10.1021/ma970617+.
37. Plazek, D. J. 1995 Bingham Medal Address: Oh, thermorheological simplicity, wherefore art thou? *Journal of Rheology (1978-present)* **1996,** 40 (6), 987-1014 DOI: doi:http://dx.doi.org/10.1122/1.550776.
38. Ninomiya, K.; Ferry, J. D.; Ōyanagi, Y. VISCOELASTIC PROPERTIES OF POLYVINYL ACETATES. II. CREEP STUDIES OF BLENDS1. *J. Phys. Chem.* **1963,** 67 (11), 2297-2308 DOI: 10.1021/j100805a012.
39. Meng, F.; Pritchard, R. H.; Terentjev, E. M. Stress Relaxation, Dynamics, and Plasticity of Transient Polymer Networks. *Macromolecules* **2016**, DOI: 10.1021/acs.macromol.5b02667.
40. Cheng, S.; Patil, S.; Cheng, S. Hydrogen Bonding Exchange and Supramolecular Dynamics of Monohydroxy Alcohols. *Phys. Rev. Lett.* **2024,** 132 (5), 058201 DOI: 10.1103/PhysRevLett.132.058201.
41. Patil, S.; Sun, R.; Cheng, S.; Cheng, S. Molecular Mechanism of the Debye Relaxation in Monohydroxy Alcohols Revealed from Rheo-Dielectric Spectroscopy. *Phys. Rev. Lett.* **2023,** 130 (9), 098201 DOI: 10.1103/PhysRevLett.130.098201.
42. Matsushima, S.; Takano, A.; Takahashi, Y.; Matsushita, Y. Dynamic viscoelasticity of a series of poly(4-n-alkylstyrene)s and their alkyl chain length dependence. *Polymer* **2017,** 133, 137-142 DOI: https://doi.org/10.1016/j.polymer.2017.10.065.
43. Rubinstein, M.; Colby, R. H., *Polymer Phyiscs*. Oxford university press: 2003.
44. Cao, Z.; Li, Z.; Zhang, S.; Galuska, L.; Li, T.; Do, C.; Xia, W.; Hong, K.; Gu, X. Decoupling Poly(3-alkylthiophenes)' Backbone and Side-Chain Conformation by Selective Deuteration and Neutron Scattering. *Macromolecules* **2020,** 53 (24), 11142-11152 DOI: 10.1021/acs.macromol.0c02086.
45. Cao, Z.; Tolba, S. A.; Li, Z.; Mason, G. T.; Wang, Y.; Do, C.; Rondeau-Gagné, S.; Xia, W.; Gu, X. Molecular Structure and Conformational Design of Donor-Acceptor Conjugated Polymers to Enable Predictable Optoelectronic Property. *Adv. Mater.* **2023,** 35 (41), 2302178 DOI: https://doi.org/10.1002/adma.202302178.
46. Cao, Z.; Leng, M.; Cao, Y.; Gu, X.; Fang, L. How rigid are conjugated non-ladder and ladder polymers? *Journal of Polymer Science* **2022,** 60 (3), 298-310 DOI: https://doi.org/10.1002/pol.20210550.

47. Root, S. E.; Savagatrup, S.; Printz, A. D.; Rodriquez, D.; Lipomi, D. J. Mechanical Properties of Organic Semiconductors for Stretchable, Highly Flexible, and Mechanically Robust Electronics. *Chem. Rev.* **2017,** 117 (9), 6467-6499 DOI: 10.1021/acs.chemrev.7b00003.
48. Li, Y.; Tatum, W. K.; Onorato, J. W.; Zhang, Y.; Luscombe, C. K. Low Elastic Modulus and High Charge Mobility of Low-Crystallinity Indacenodithiophene-Based Semiconducting Polymers for Potential Applications in Stretchable Electronics. *Macromolecules* **2018,** 51 (16), 6352-6358 DOI: 10.1021/acs.macromol.8b00898.
49. Sirringhaus, H.; Brown, P. J.; Friend, R. H.; Nielsen, M. M.; Bechgaard, K.; Langeveld-Voss, B. M. W.; Spiering, A. J. H.; Janssen, R. A. J.; Meijer, E. W.; Herwig, P.; de Leeuw, D. M. Two-dimensional charge transport in self-organized, high-mobility conjugated polymers. *Nature* **1999,** 401 (6754), 685-688 DOI: 10.1038/44359.
50. Noriega, R.; Rivnay, J.; Vandewal, K.; Koch, F. P. V.; Stingelin, N.; Smith, P.; Toney, M. F.; Salleo, A. A general relationship between disorder, aggregation and charge transport in conjugated polymers. *Nat. Mater.* **2013,** 12 (11), 1038-1044 DOI: 10.1038/nmat3722.
51. Gu, K.; Loo, Y.-L. The Polymer Physics of Multiscale Charge Transport in Conjugated Systems. *J. Polym. Sci., Part B: Polym. Phys.* **2019,** 57 (23), 1559-1571 DOI: https://doi.org/10.1002/polb.24873.
52. Fratini, S.; Nikolka, M.; Salleo, A.; Schweicher, G.; Sirringhaus, H. Charge transport in high-mobility conjugated polymers and molecular semiconductors. *Nat. Mater.* **2020,** 19 (5), 491-502 DOI: 10.1038/s41563-020-0647-2.
53. Wagner, H. L. The Mark–Houwink–Sakurada Equation for the Viscosity of Atactic Polystyrene. *J. Phys. Chem. Ref. Data* **1985,** 14 (4), 1101-1106 DOI: 10.1063/1.555740.
54. Bushin, S. V.; Bezrukova, M. A.; Koeckelberghs, G.; Persoons, A.; Yakimansky, A. V. Model of a polymer chain twisted inside a statistical segment. Poly(3-hexylthiophene) in chloroform solutions. *Polym. Int.* **2017,** 66 (6), 869-875 DOI: https://doi.org/10.1002/pi.5326.

**Acknowledgments:** This research was performed on APS beam time award(s) (DOI: https://doi.org/10.46936/APS-191684/60015416) from the Advanced Photon Source, a U.S. Department of Energy (DOE) Office of Science user facility operated for the DOE Office of Science by Argonne National Laboratory under Contract No. DE-AC02-06CH11357. The team thanks Vanessa Phan and Graham Collier for helping with SEC test.

**Funding:** S.C. acknowledges the support of National Science Foundation with the Award Number NSF-DMR 2211573. X.G. thanks Office of Naval Research for support this project under grant number N-00014-23-1-2063.

**Author contributions:** S.C. conceived the research idea. S.C. and X.G. supervise the research. J.Z. performed all rheological measurements. J.Z. and X.Z. conducted the X-ray measurements. J.Z., S.C., and X.G. performed the data analysis. J.Z. wrote the initial draft of the manuscript. All authors reviewed and edited the manuscript.

**Competing interests:** The authors declare that they have no competing interests.

**Data and materials availability:** All data needed to evaluate the conclusions in the paper are present in the paper and/or the Supplementary Materials. Raw data are available from the corresponding authors upon request.

**SUPPLEMENTARY MATERIALS**

# Reversible Interchain Associations Dominate Plateau Elasticity in Conjugated Polymers

Juncheng Zheng,[1] Xiaobing Zuo,[2] Xiaodan Gu,[3,*] and Shiwang Cheng[1,*]

[1]Department of Chemical Engineering and Materials Science, Michigan State University, East Lansing, Michigan 48824, United States
[2]X-ray Science Division, Argonne National Laboratory, Lemont, Illinois 60439, United States
[3]School of Polymer Science and Engineering, Center for Optoelectronic Materials and Devices, The University of Southern Mississippi, Hattiesburg, Mississippi 39406, United States

**The file includes:**

Figures S1-S2: SEC characterization and analysis;
Figure S3: DSC for P3Ats;
Figure S4: Temperature-dependent SAXS/WAXS of P3BT and P3OT;
Figure S5: Dynamic shift factors and modulus shift factors of P3HT and PS;
Figure S6: Temperature-dependent creep and creep recovery of P3HT and PS;
Figure S7: Temperature dependence of characteristic relaxation times in PS;
Figure S8: Creep and creep recovery of PS under varying stress;
Figure S9: Strain sweep of P3HT at $T = 328\ K$ and PS at $T = 403\ K$ under $\omega = 10\ s^{-1}$;
Figure S10: Sequential SAOS/MAOS/SAOS measurement on PS at $T = 403\ K$ ;
Figure S11: Sequential SAOS/MAOS/SAOS measurement on P3HT at $T = 373\ K$ ;
Figure S12: Master curves, Dynamic shift factors, and van-Gurp plots of P3BT and P3OT;
Figure S13: Temperature-dependent creep response of P3BT and P3OT.

* Corresponding Authors. Email Address: xiaodan.gu@usm.edu (X.G) and chengsh9@msu.edu (S.C.)

## Supplementary Figures

### Size exclusion chromatography (SEC) characterization and analysis

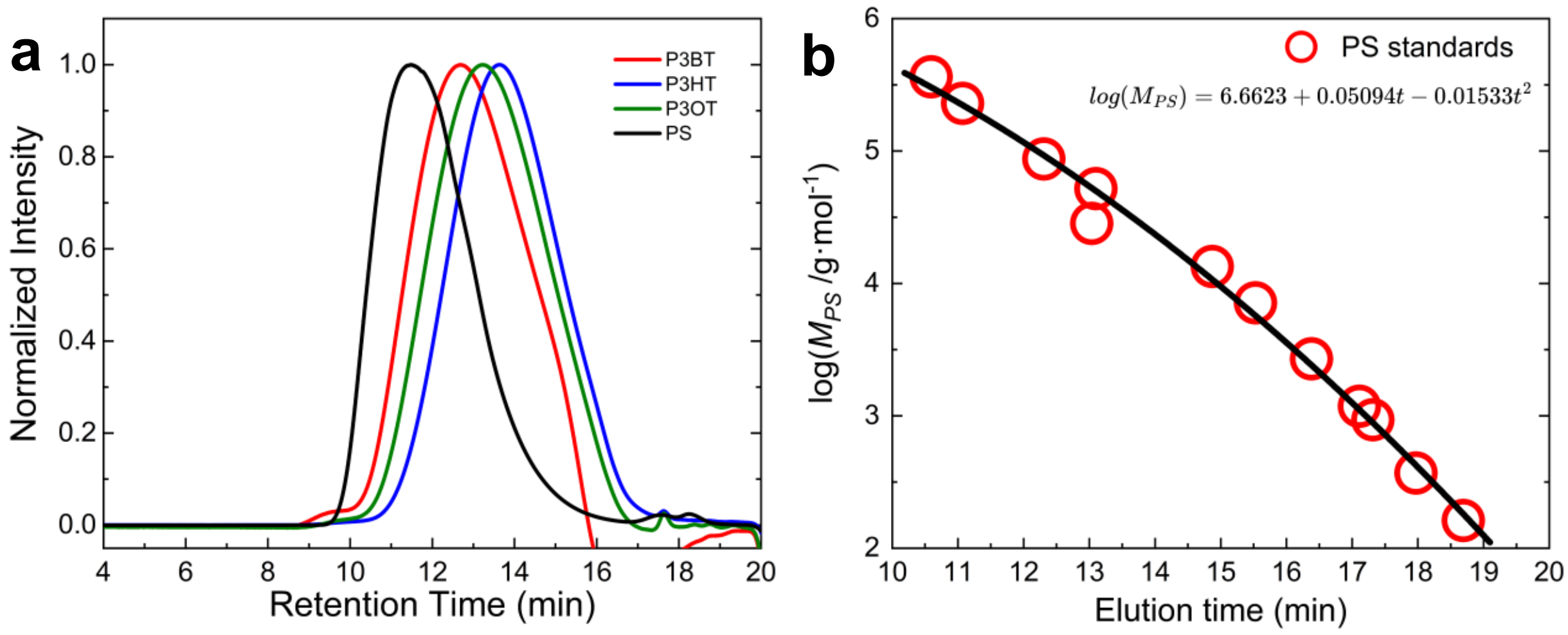

**Figure S1. (a)** SEC traces of P3BT (red), P3HT (blue), P3OT (green), and PS (black). **(b)** The calibration curve of SEC using PS standards with molecular weights ranging from 162 to 364,000 g/mol, from which a universal calibration curve can be obtained to determine the molecular weight of the conjugated polymers.

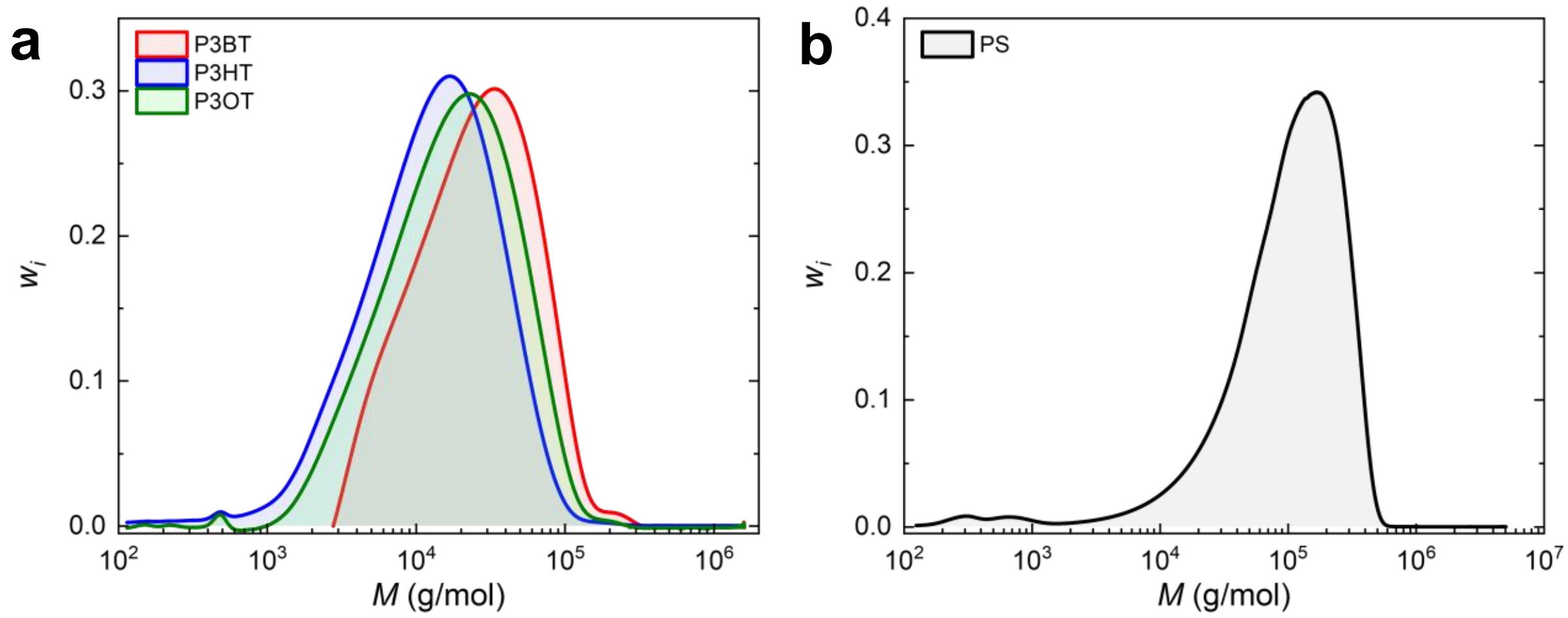

**Figure S2. (a-b)** Weight fraction distribution of molecular weight for **(a)** P3BT (red), P3HT (blue), and P3OT (green), and **(b)** PS (black).

**Differential scanning calorimetry (DSC)**

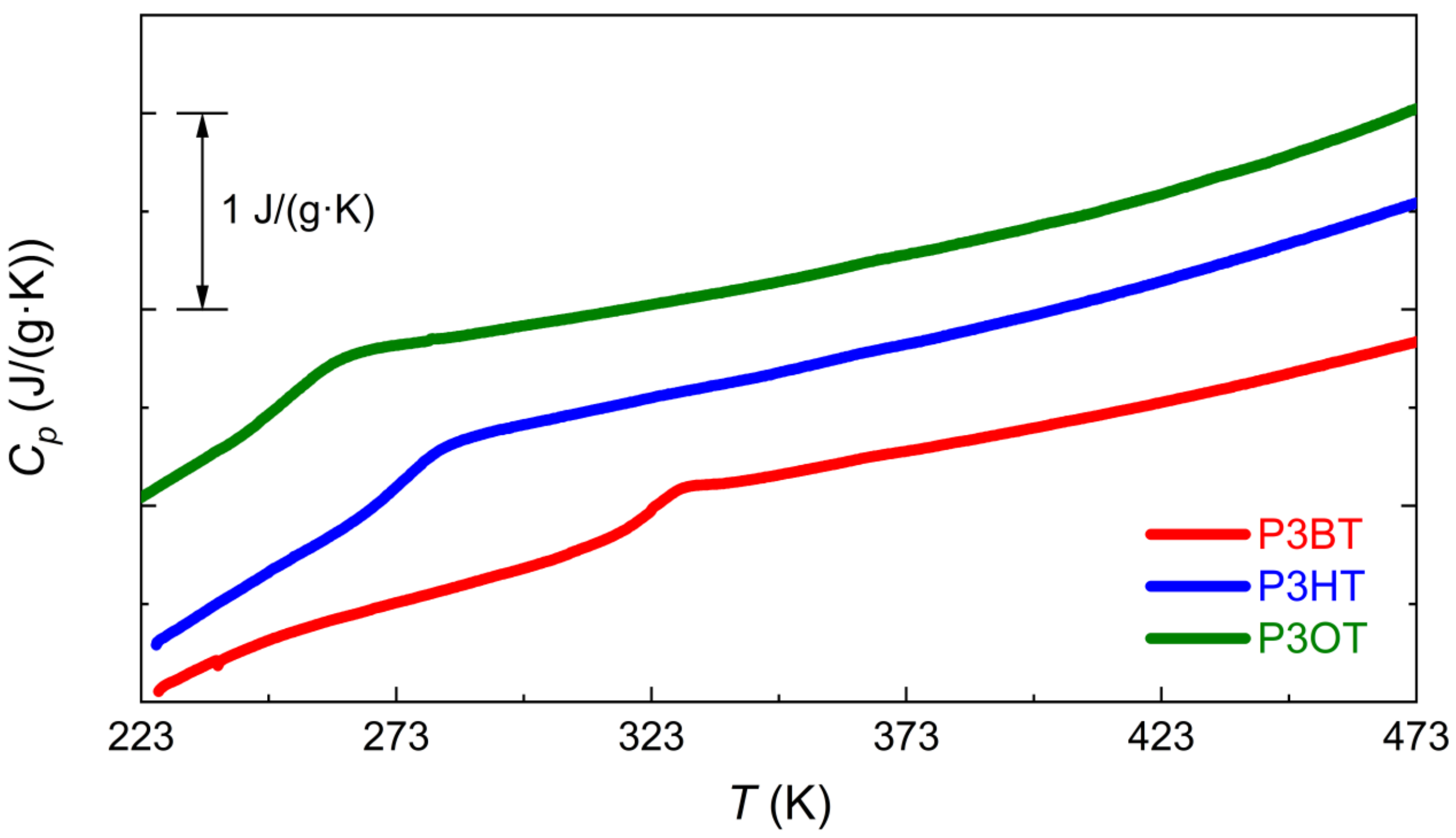

**Figure S3.** Specific heat capacity change, $C_p$, upon heating (second heating) at a rate of $10\ K/min$ for P3BT (red), P3HT (blue), and P3OT (green).

**Small- and wide-angle X-ray scattering (SAXS/WAXS) of P3BT and P3OT**

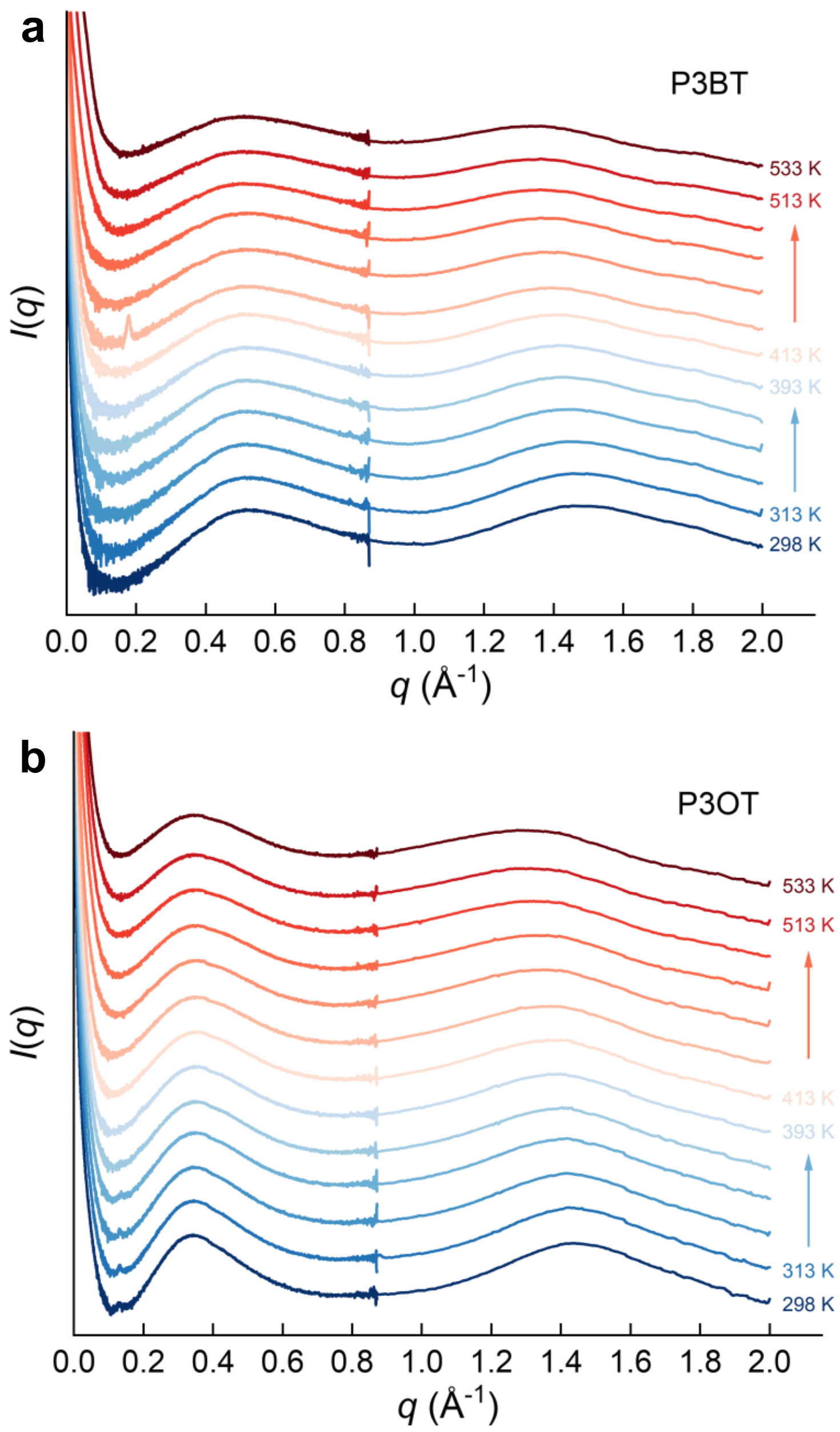

**Figure S4.** SAXS/WAXS measurements of **(a)** P3BT and **(b)** P3OT at the temperature of $298\ K - 533\ K$ at an interval of $20\ K$. No signs of crystalline order or liquid crystalline order were found in both P3BT and P3OT.

**Dynamic shift factors, $a_T^S$, and modulus shift factors, $b_T$, of P3HT and PS**

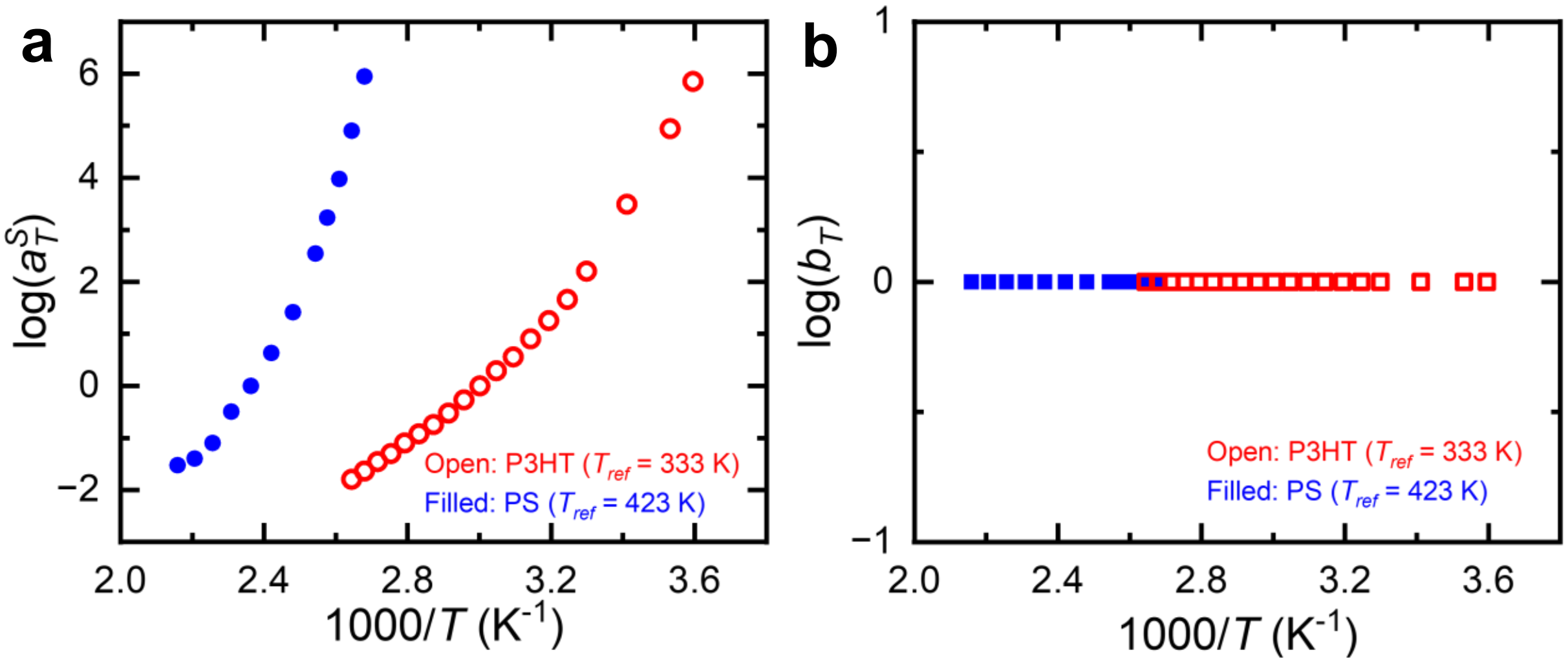

**Figure S5. (a)** Dynamic shift factors, $a_T^S$, of P3HT (red open circle) at $T_{ref} = 333\ K$ and PS (blue filled circle) at $T_{ref} = 423\ K$. **(b)** Modulus shift factors, $b_T$, of P3HT (red open square) at $T_{ref} = 333\ K$ and PS (blue filled square) at $T_{ref} = 423\ K$.

**Temperature-dependent creep and creep recovery of P3HT and PS**

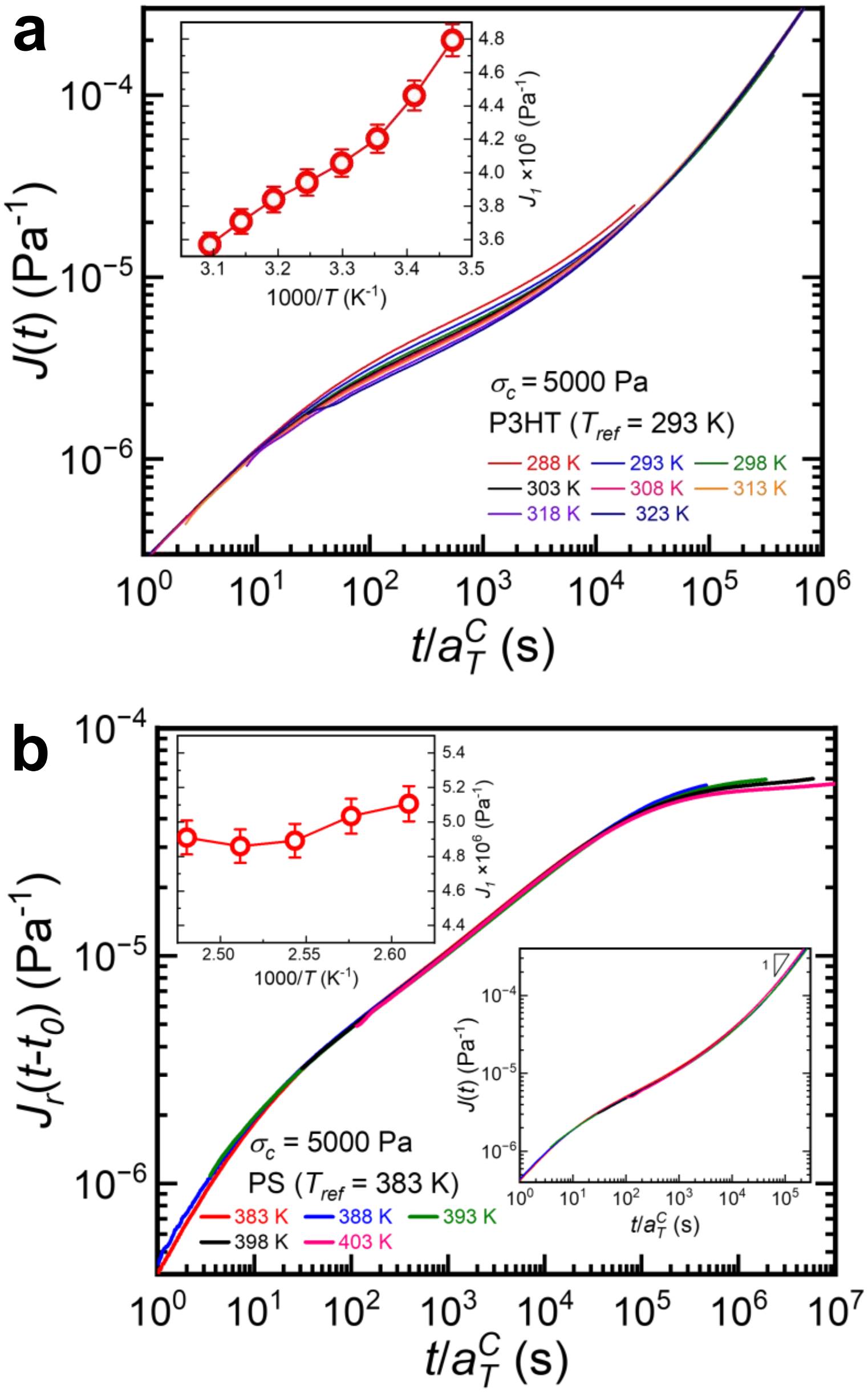

**Figure S6. (a)** Creep compliance, $J(t)$, of P3HT at $\sigma_C = 5{,}000\ Pa$ at $288\ K - 323\ K$ at $T_{ref} = 293\ K$. The inset shows the temperature dependence of intermediate plateau compliance, $J_1$, upon creep. **(b)** Recoverable creep compliance, $J_r(t - t_0)$, of PS after a creep at $\sigma_C = 5{,}000\ Pa$ at $383\ K - 403\ K$ and $T_{ref} = 383\ K$. The lower inset presents the corresponding creep compliance, $J(t)$. The upper inset shows the temperature dependence of the plateau compliance, $J_1$.

**Temperature dependence of characteristic relaxation times in PS**

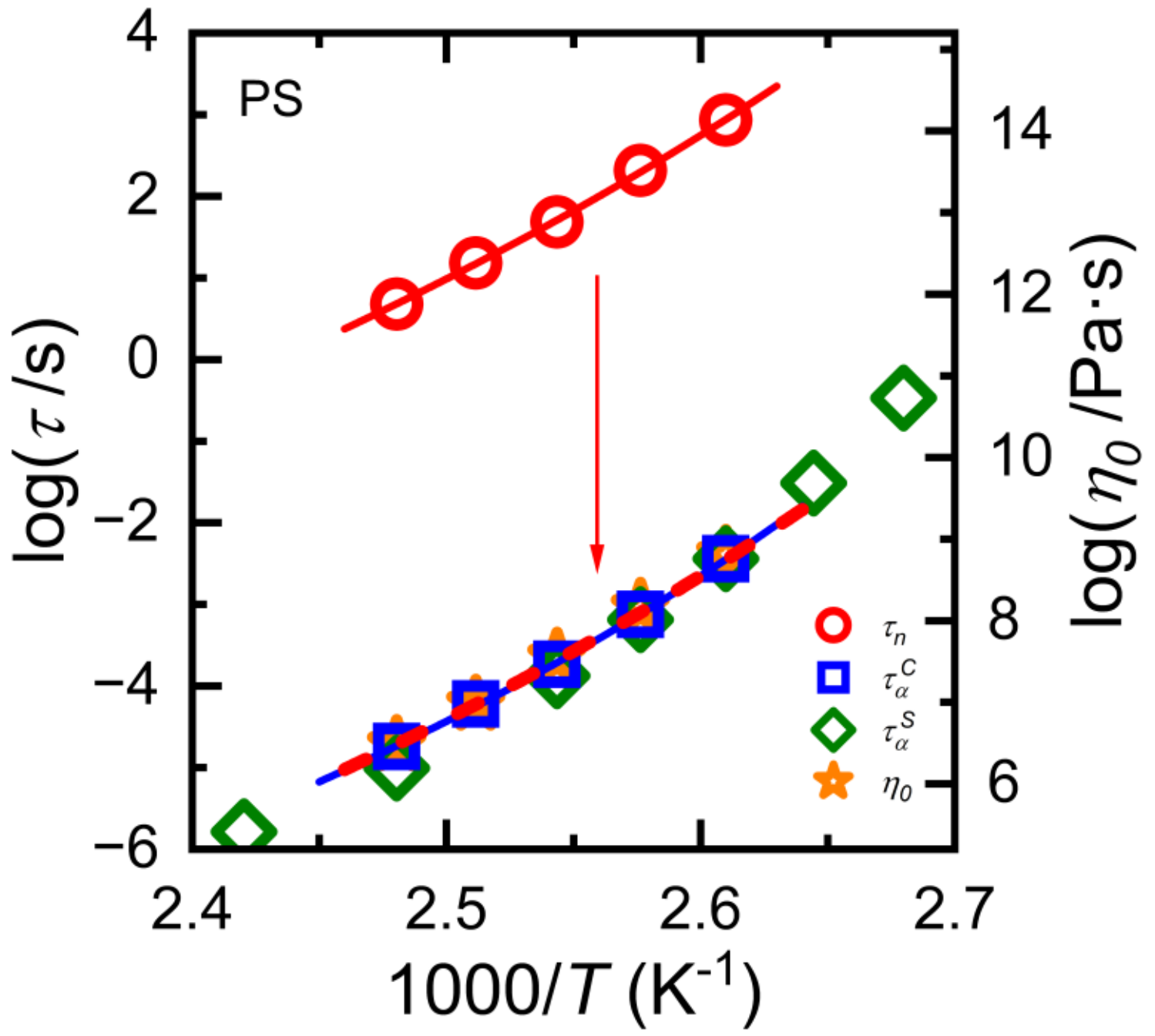

**Figure S7.** The activation plot of the characteristic time of modulus plateau relaxation (red circles), $\tau_n$, the segmental relaxation times from creep measurement (blue square), $\tau_\alpha^C$, the segmental relaxation time from SAOS measurements (green diamond), $\tau_\alpha^S$, and zero-shear viscosity (orange star), $\eta_0$, of PS. The red solid line presents the Vogel-Fulcher-Tamman (VFT) fit to $\tau_n$, and the blue solid line presents the VFT fit to $\tau_\alpha^C$. Identical temperature dependences of $\tau_n$, $\tau_\alpha^C$, $\tau_\alpha^S$, and $\eta_0$ are observed.

## Creep and creep recovery of PS under varying stress

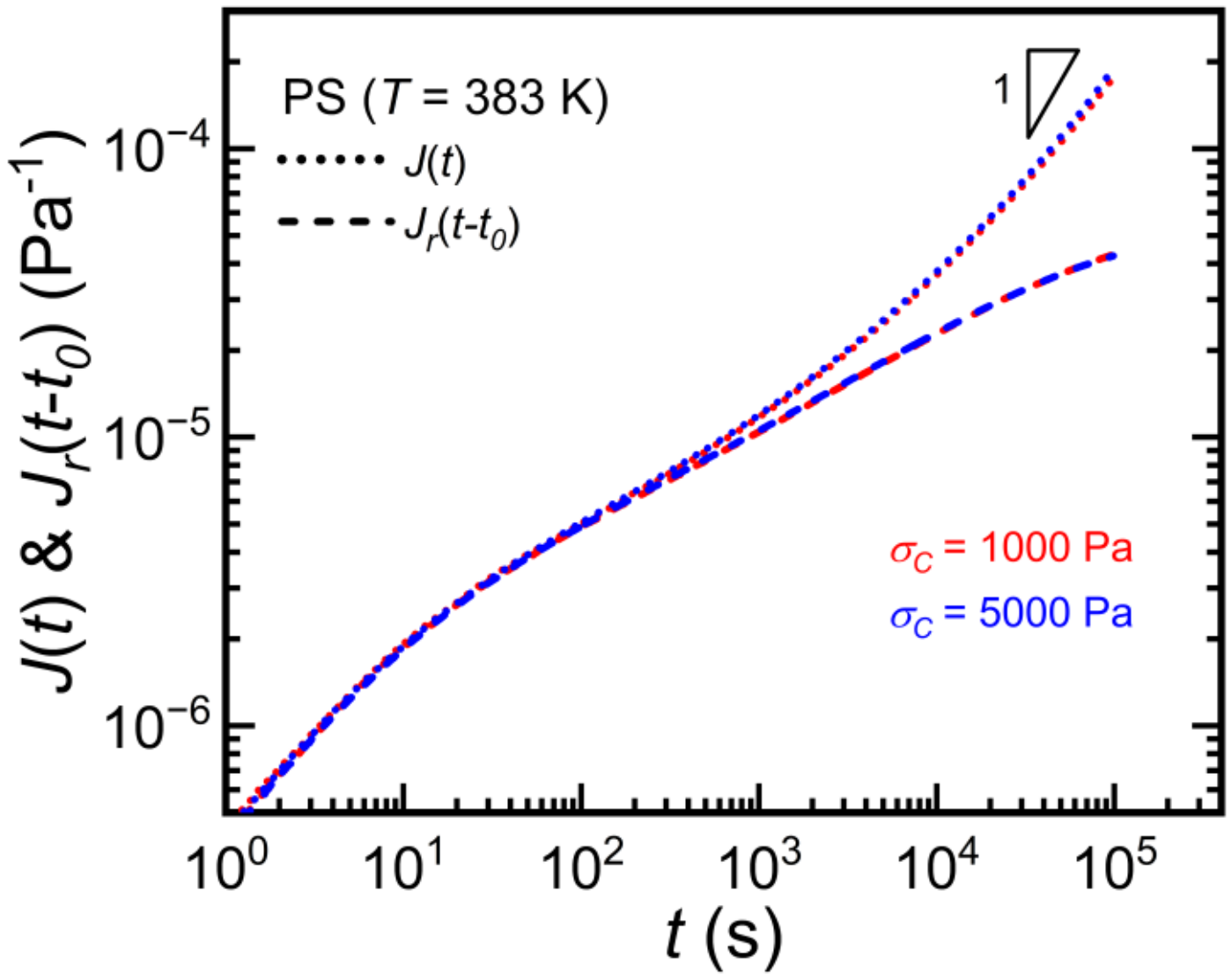

**Figure S8.** Creep compliance (dotted line), $J(t)$, and recoverable creep compliance (dashed line), $J_r(t-t_0)$, of PS at $383\,K$ under different creep stresses, $\sigma_C = 1{,}000\,Pa$ (red) and $\sigma_C = 5{,}000\,Pa$ (blue).

**Strain sweep of P3HT at $T = 328\ K$ and PS at $T = 403\ K$ under $\omega = 10\ s^{-1}$**

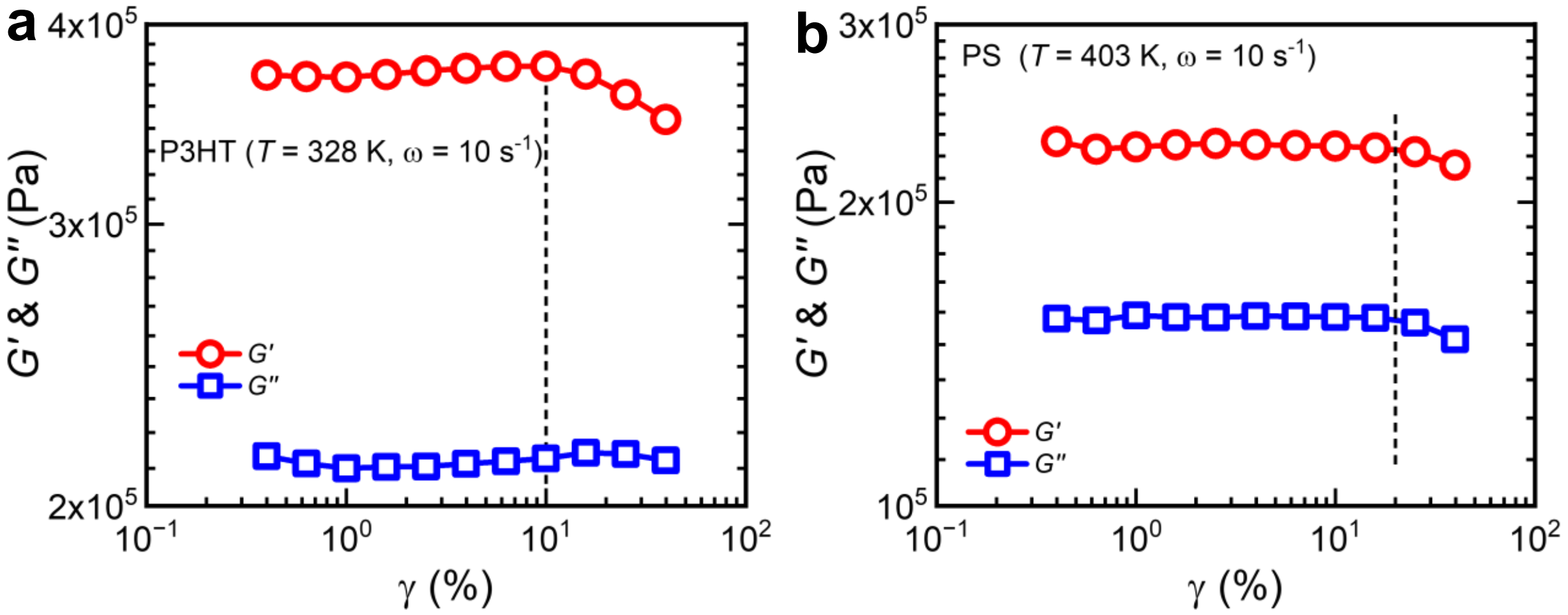

**Figure S9. (a-b)** Storage moduli, $G'$ (red circle), and loss moduli, $G''$ (blue square), as a function of oscillatory shear strain amplitude, $\gamma$, at a constant angular frequency $\omega = 10\ rad * s^{-1}$ for **(a)** P3HT at $T = 328\ K$ and **(b)** PS at $T = 403\ K$. The dashed line indicates the end of the linear response regime, which provides the oscillatory strain amplitude for the MAOS measurements.

**Sequential SAOS/MAOS/SAOS measurement on PS at $T = 403\ K$**

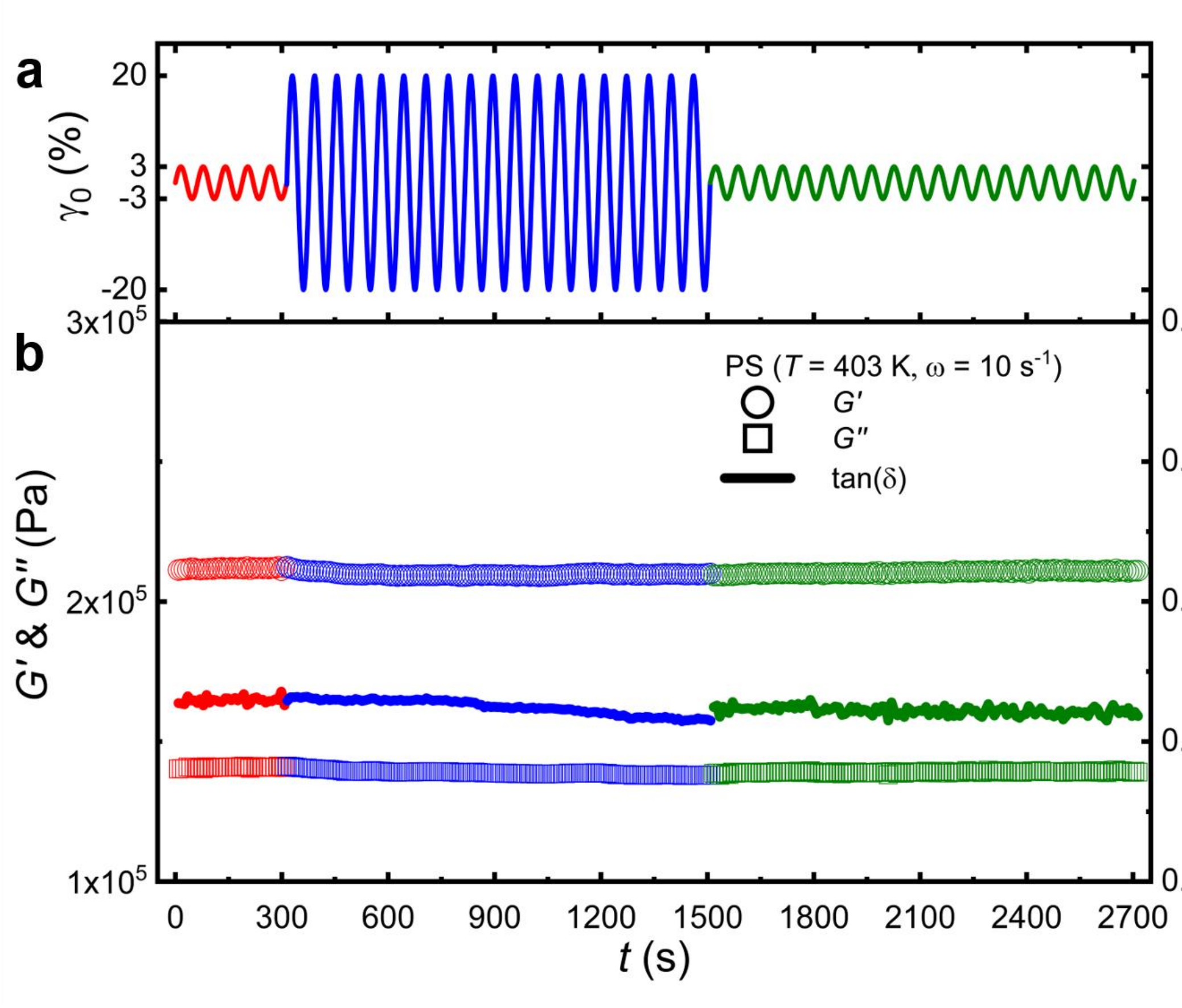

**Figure S10. (a)** SAOS/MAOS/SAOS measurements on PS at $403\ K$ at a constant angular frequency $\omega = 10\ s^{-1}$. A strain amplitude of $\gamma_0 = 3\%$ is applied to the initial and final SAOS stages and a $\gamma_0 = 20\%$ is employed in the middle MAOS stage. **(b)** Time-dependent dynamic moduli, $G'$ (circle) and $G''$ (square), and the loss factor, $tan(\delta)$ (solid line), of PS upon SAOS/MAOS/SAOS measurement corresponding to the protocol in (a).

**Sequential SAOS/MAOS/SAOS measurement on P3HT at $T = 373\ K$**

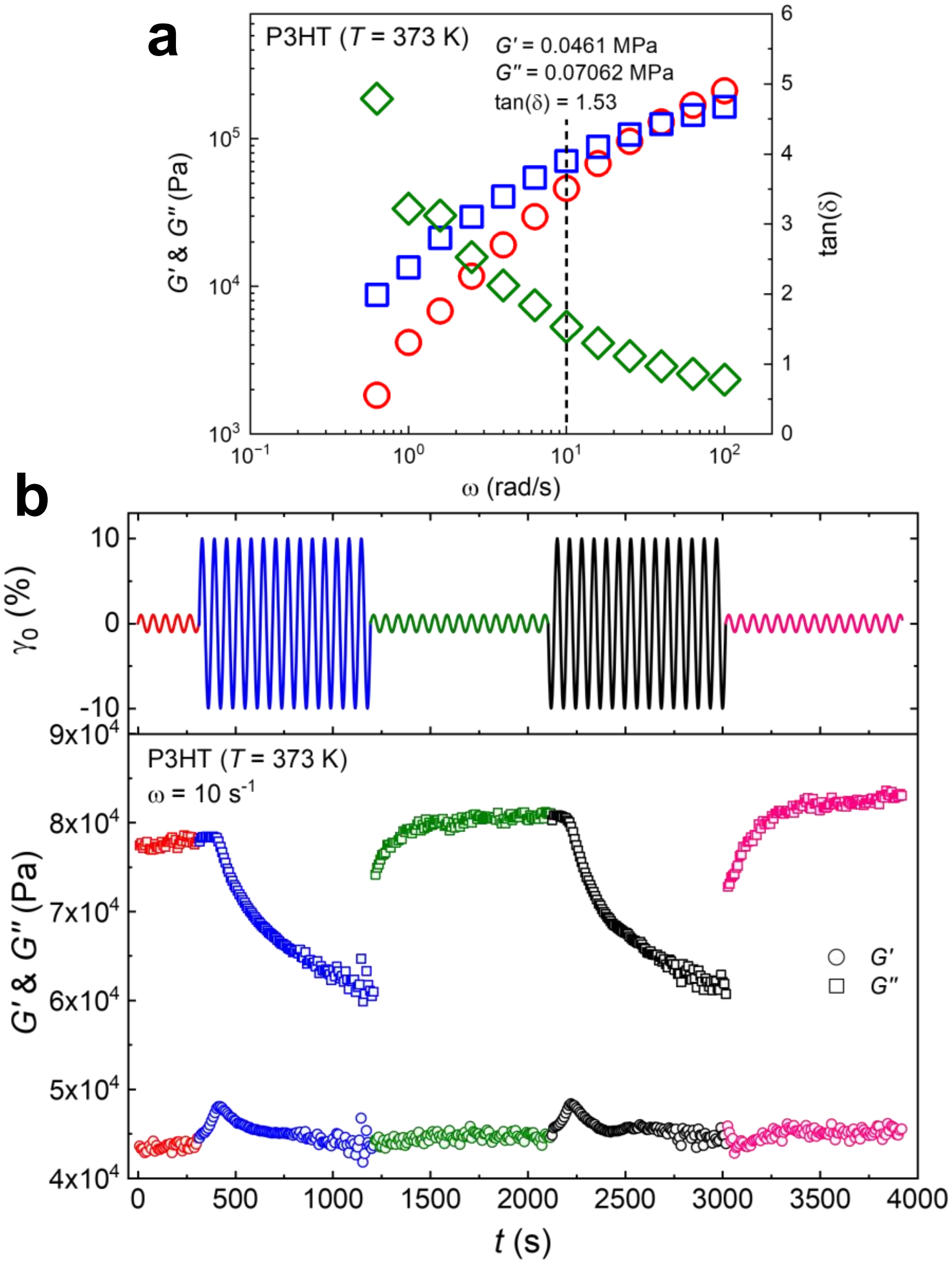

**Figure S11. (a)** Storage moduli, $G'$ (red circles), loss moduli, $G''$ (blue squares), and loss factor, $tan\delta$, as a function of angular frequency, $\omega$, for P3HT at $T = 373K$. **(b)** Time-dependent moduli ($G'$ (circles) and $G''$ (squares)) under sequential SAOS/MAOS/SAOS/MAOS/SAOS at $T = 373K$.

**Master curves, Dynamic shift factors, and van-Gurp plots of P3BT and P3OT**

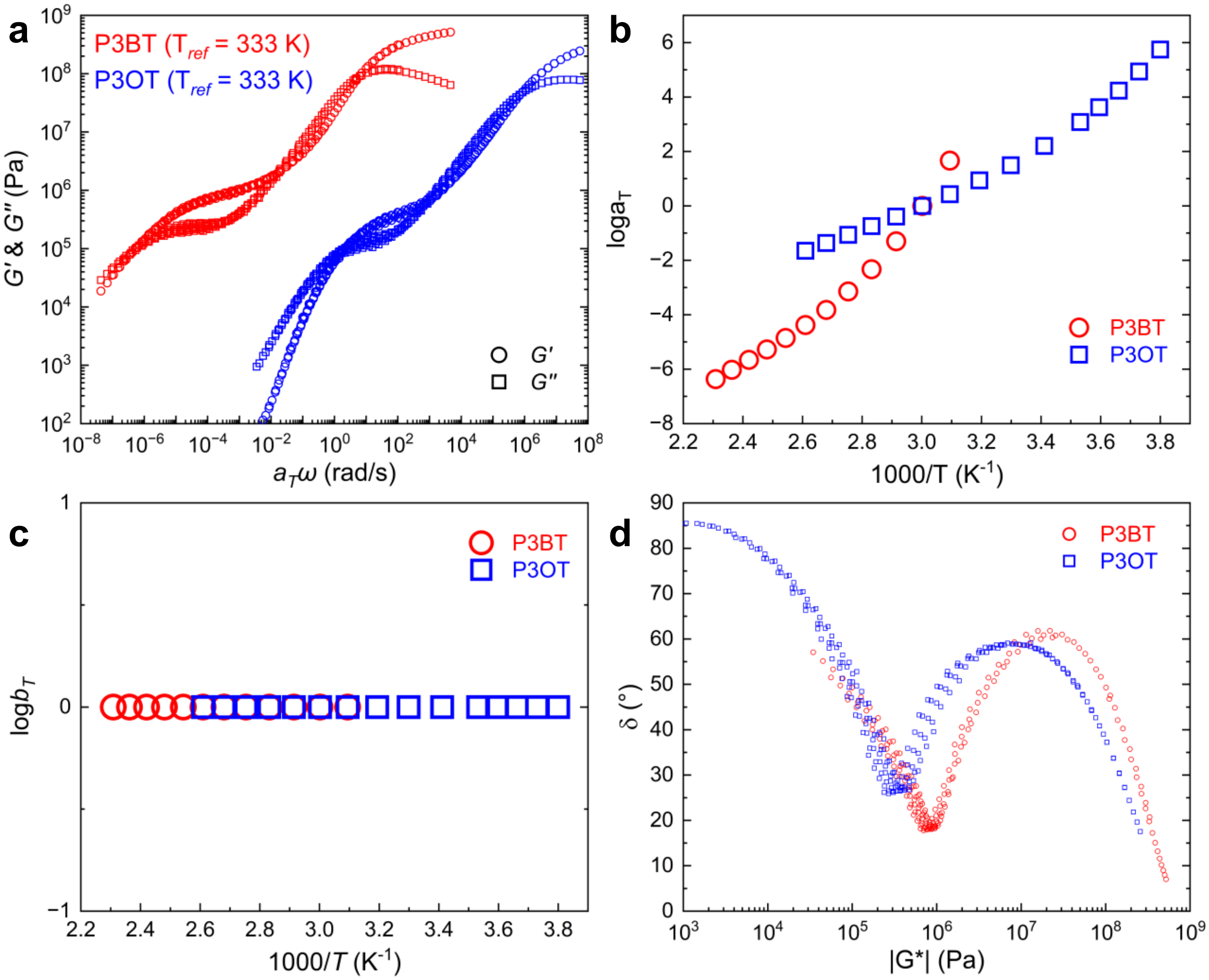

**Figure S12. (a)** Linear viscoelasticity master curves of P3BT (red) and P3OT (blue) at a reference temperature of $T_{ref} = 333\ K$. **(b)** Dynamic shift factors, $a_T^S$, for P3BT (red circles) and P3OT (blue squares) at $T_{ref} = 333\ K$. **(c)** The corresponding modulus shift factors, $b_T$, for P3BT (red circles) and P3OT (blue squares) at $T_{ref} = 333\ K$. **(d)** van Gurp-Palmen plot, the phase angle $\delta$ versus absolute complex modulus $|G^*|$, for P3BT (red circle) and P3OT (blue square). For both P3BT and P3OT, significant deviations for the spectra at different testing temperatures were observed at the $|G^*| \sim 10^5 - 10^6$ Pa.

**Temperature-dependent creep response of P3BT and P3OT**

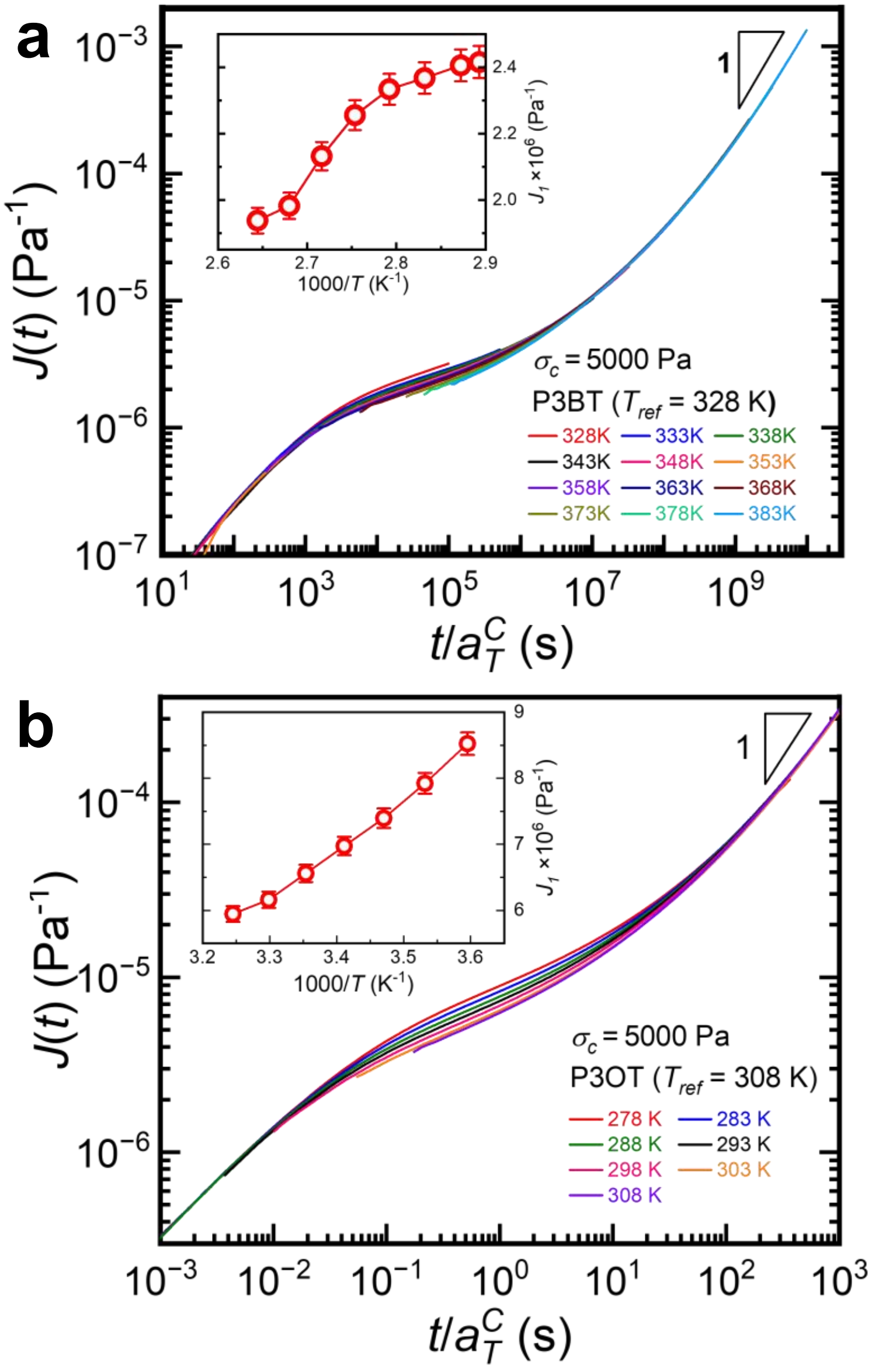

**Figure S13. (a)** Creep compliance, $J(t)$, of P3BT at $\sigma_C = 5{,}000\ Pa$, $T = 328\ K - 383\ K$, and $T_{ref} = 328\ K$. The inset shows the corresponding, $J_1$, at each test temperature. **(b)** Creep compliance, $J(t)$, of P3OT at $\sigma_C = 5000\ Pa$, $T = 278\ K - 308\ K$, and $T_{ref} = 308\ K$. The inset shows the corresponding, $J_1$, at each temperature.